\documentclass[preprint,superscriptaddress,showpacs,preprintnumbers,amsmath,amssymb]{revtex4}

\usepackage{graphicx}
\usepackage{dcolumn}
\usepackage{bm}

\begin{document}

\title{Supercurrent induced by tunneling Bogoliubov excitations in a Bose-Einstein condensate}

\author{Shunji Tsuchiya}
\author{Yoji Ohashi}%
\affiliation{Department of Physics, Keio University, 3-14-1 Hiyoshi,
Kohoku-ku, Yokohama 223-8522, Japan}
\affiliation{CREST-JST, 4-1-8 Honcho, Saitama 332-0012, Japan}

\date{\today}

\begin{abstract}
We study the tunneling of Bogoliubov excitations through a barrier in a
Bose-Einstein condensate.
We extend our previous work [Phys. Rev. A \textbf{78}, 013628 (2008)]
 to the case when condensate densities are different between the left
 and right of the barrier potential.
In the framework of the Bogoliubov mean-field theory, we calculate the
transmission probability and phase shift, as well as the energy flux
and quasiparticle current carried by Bogoliubov excitations.
We find that Bogoliubov phonons twist the condensate phase
due to a back-reaction effect, which induces the Josephson
 supercurrent.
While the total current given by the sum of quasiparticle current and
 induced supercurrent is conserved, the quasiparticle current
 flowing through the barrier potential is
 shown to be remarkably enhanced in the low energy region. 
When the condensate densities are different between the left and right
 of the barrier, the excess quasiparticle current, as well as the
 induced supercurrent, remains finite far
 away from the barrier.
We also consider the tunneling of excitations and atoms through the boundary
 between the normal and superfluid regions.
We show that supercurrent can be generated inside the condensate by injecting
free atoms from outside. On the other hand, atoms are emitted when the
 Bogoliubov phonons propagate toward the phase boundary from the
 superfluid region.

\end{abstract}

\maketitle

\section{introduction}

In the pioneering work by Bogoliubov \cite{Bogoliubov}, it was
shown that the Bose-Einstein condensate (BEC) of weakly interacting bosons
has a phonon-type excitation mode. 
It is now called the Bogoliubov mode, which is a Nambu-Goldstone mode
associated with a spontaneous broken U(1) symmetry \cite{Anderson}. This
collective mode dominates low-energy properties of BEC, so that 
it is an important key to understand physical properties of BEC \cite{Pitaevskii}.
In particular, the existence of Bogoliubov phonon is essential for the Bose-condensed phase to acquire superfluidity \cite{Landau}.
Since the realization of BECs in ultracold atomic gases \cite{M.H.Anderson,Davis},
the study of Bogoliubov mode has been one of the main issues in cold
atom physics \cite{Pitaevskii,Davidson}. 
Because of the high degree of controllability, the BECs of cold atomic gases
offer good opportunities to explore novel properties of Bogoliubov excitations.

Recently, Kovrizhin and co-workers \cite{Kovrizhin1,Kovrizhin2,Kagan} predicted that
the Bogoliubov mode exhibits striking tunneling properties.
They showed that the transmission probability of Bogoliubov
phonon through a potential barrier increases in the low energy region with decreasing the incident energy.
In the low-energy limit, the perfect transmission is realized
irrespective of the height of the barrier. 
This interesting tunneling property of Bogoliubov mode is referred to as
the {\it anomalous tunneling} \cite{Kagan}. 
Since their prediction \cite{Kovrizhin1,Kovrizhin2,Kagan}, the anomalous
tunneling has attracted much attention, and has been addressed by many papers
\cite{Danshita1,Danshita2,Danshita3,Bilas,Kato,Tsuchiya,Ohashi,Watabe,Fujita}. 

As the origin of the anomalous tunneling effect, various mechanisms have
been proposed, such as quasiresonance scattering \cite{Kagan},
localized components of Bogoliubov mode appearing near the barrier \cite{Danshita1},
and anomalous enhancement of quasiparticle current \cite{Tsuchiya}.
For the perfect transmission in the
low-energy limit, the importance of the coincidence of the condensate and excitation
wave functions \cite{Kato}, as well as supercurrent behavior of low-energy 
Bogoliubov phonons \cite{Ohashi}, has been pointed out.
The anomalous tunneling phenomenon was shown to occur even in the
supercurrent state \cite{Danshita1,Ohashi,Takahashi}, as well as at finite temperatures \cite{Kato}.
It has been also studied in the presence of a periodic potential
\cite{Danshita2,Danshita3}, as well as a random potential \cite{Bilas}.
It has been also pointed out that similar phenomena to this can be seen
in the scattering of Bogoliubov phonon by a spherical potential in three
dimensions d\cite{Fujita},
as well as the refraction of Bogoliubov phonons \cite{Watabe}.

In this paper, we investigate tunneling properties of Bogoliubov phonon
in a BEC at $T=0$. In Ref.~\cite{Tsuchiya}, we have considered the case
when the incident and transmitted Bogoliubov phonons feel the same
condensate densities on both the right and left of the barrier. 
In this paper, we extend this previous paper to the
case when the condensate density is different between the right and the
left of the barrier. As an extreme case, we also deal with the case when
the condensate density is absent on one side of the barrier. 
Applying the finite element method to the Bogoliubov coupled equations, we
numerically calculate the transmission probability and phase shift of
Bogoliubov phonons.
We find that Bogoliubov phonons twist the phase of the BEC order
parameter (condensate wave function) due to a back-reaction effect,
which leads to the induction of Josephson supercurrent.
The induced supercurrent is shown to satisfy the Josephson
relation with respect to the twisted phase when the condensate density
is the same on both sides of the barrier.
The supercurrent is induced only in the region near the barrier when the condensate
has the same densities across the potential barrier. In the case when
the condensate density is different between the right and left of 
the barrier, the supercurrent is also induced in the region far away from the barrier.
In addition, the excess quasiparticle current is supplied from the condensate to
conserve the total current, so that one obtains the enhancement of the transmission
probability of quasiparticle current in the low-energy region.
We also show that the supercurrent is induced when one injects free
atoms from the outside of condensate. In addition, atoms are shown to evaporate from the
surface of superfluid region when Bogoliubov excitation propagates
toward the superfluid-normal phase boundary.

This paper is organized as follows:
in Sec.~\ref{sec2}, we present the model and formalism of the Bogoliubov
mean-field approximation, as well as the finite element method which we apply
for solving the Bogoliubov equations.
In Sec.~\ref{sec3}, we study the tunneling of Bogoliubov phonons
through a rectangular potential barrier. We give a detailed discussion on
the origin of the anomalous tunneling and induced Josephson supercurrent.
In Sec.~\ref{sec4}, we study the tunneling in the presence of a step
potential which yields the different condensate densities across the
potential barrier. 
In Sec.~\ref{sec5}, we discuss the tunneling of excitations and atoms
between superfluid and normal regions.

\section{model and formalism}
\label{sec2}

We consider tunneling phenomena of Bogoliubov mode through a barrier
potential, as schematically shown in Fig.~1. We assume that the barrier
potential $U(x)$ only depends on $x$ and ignore the motion of atoms in
the $y$ and $z$ directions, so that we consider a
one-dimensional tunneling problem along the $x$ direction.
This kind of one-dimensional geometry has been recently realized
 \cite{Engels}. In Ref.~\cite{Engels}, a BEC was prepared in a narrow
elongated trap with a wall-type potential barrier, which varies only in the axial
direction and the potential width is much longer than the radial size of
 the gas cloud.
We also ignore temperature effects as well as effects of a harmonic trap. 
The latter assumption is justified when the BEC is trapped in an elongated trap
\cite{Engels,Andrews} or a box-shaped trap \cite{Meyrath}.

\begin{figure}
\centerline{\includegraphics{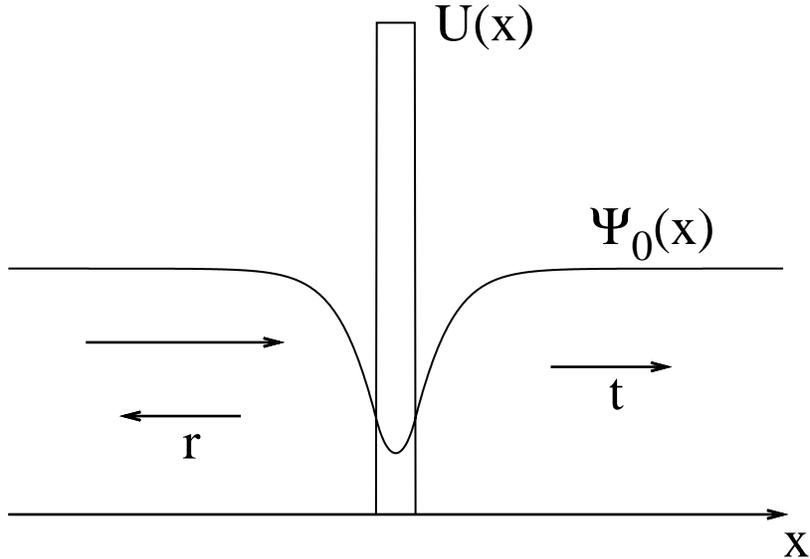}}
\caption{Schematic of the system. The arrows on the left
 describe the incoming (upper arrow) and reflected (lower arrow)
 Bogoliubov excitations. The arrow on the right describes the transmitted
 one.}
\label{system}
\end{figure}

We treat the tunneling of Bogoliubov mode within the Bogoliubov
mean-field theory for a weakly interacting
Bose gas at $T=0$ \cite{Bogoliubov,Pitaevskii2,Fetter}.
To describe the BEC phase, we divide the Bose 
field operator $\hat\psi(\bm r)$ into the condensate wave function $\Psi_0(\bm r)$ and the
noncondensate part, as
\begin{equation}
\hat\psi(\bm r)=\Psi_0(\bm r)+\sum_j\left[u_j(\bm r)\hat\alpha_j-v_j(\bm
				     r)^\ast\hat\alpha_j^\dagger\right],
\label{Bogoansatz}
\end{equation}
where $\hat\alpha_j^\dagger$ is the creation operator of a Bogoliubov
excitation in the $j$th state.
The condensate wave function $\Psi_0(\bm r)=\langle\hat\psi(\bm r)\rangle$
satisfies the static Gross-Pitaevskii (GP) equation \cite{Pitaevskii2,Gross},
\begin{eqnarray}
\left(-\frac{\nabla^2}{2m}+U(\bm r)+g|\Psi_0|^2\right)\Psi_0=\mu
 \Psi_0.
\label{GP}
\end{eqnarray}
Here, $m$, $\mu$, and $U(\bm r)$ represent the mass of a boson, chemical
potential, and barrier potential, respectively.
$g(>0)$ is a repulsive interaction between bosons.
In Eq.~(\ref{Bogoansatz}), $u_j(\bm r)$ and $v_j(\bm r)$ satisfy
the Bogoliubov coupled equations,
\begin{eqnarray}
\left[-\frac{\nabla^2}{2m}+U(\bm r)+2g|\Psi_0|^2-\mu\right]u_j&\!-\!&g\Psi_0^2v_j=E_ju_j,\label{Bogoliubovu}\\
\left[-\frac{\nabla^2}{2m}+U(\bm r)+2g|\Psi_0|^2-\mu\right]v_j&\!-\!&g(\Psi_0^\ast)^2u_j=-E_jv_j,
\label{Bogoliubovv}
\end{eqnarray}
where $E_j$ is the Bogoliubov excitation spectrum.
To solve Eqs.~(\ref{Bogoliubovu}) and (\ref{Bogoliubovv}) with an appropriate boundary condition, 
we use the finite element method \cite{FEM}.
For this purpose, it is convenient to rewrite Eqs.~(\ref{Bogoliubovu})
and (\ref{Bogoliubovv}) in the matrix form
\begin{equation}
\left(\hat{\bar{H}}-\bar E_j\tau_3\right)\phi_j=0,
\label{Bogomatrix}
\end{equation} 
where
\begin{equation}
\hat{\bar H}=
\left(
\begin{matrix}
-\bar{\nabla}^2+\bar{U}(\bar{\bm r})+2|\bar{\Psi}_0|^2-1 & -\bar{\Psi}_0^2 \\
-(\bar{\Psi}_0^\ast)^2 & -\bar{\nabla}^2+\bar{U}(\bar{\bm r})+2|\bar{\Psi}_0|^2 -1
\end{matrix}
\right),
\end{equation}
\begin{equation}
\phi_j=
\left(
\begin{array}{l}
u_j(\bar{\bm r})\\
v_j(\bar{\bm r})
\end{array}
\right),
\end{equation}
\begin{equation}
\tau_3=\left(\begin{matrix}1&0\\0&-1\end{matrix}\right).
\end{equation}
In Eq.~(\ref{Bogomatrix}), we have introduced dimensionless variables $\bar{\bm
r}\equiv\bm r/\xi$, $\bar{E}_j\equiv E_j/\mu$, $\bar{U}\equiv U/\mu$, and
$\bar{\Psi}_0\equiv\Psi_0/n_0$, where $n_0\equiv\mu/g$ is the condensate
density far away from the barrier, and $\xi\equiv 1/\sqrt{2mgn_0}$ is the healing length.
To simplify our notations, we omit the bars and indices of eigenstates
in the following part of this section.
Equation~(\ref{Bogomatrix}) can be obtained from the variational
principle $\delta L=0$, when the Lagrangian $L$ has the form
\begin{eqnarray}
L=
\int_\Omega d\bm
 r\left[(\nabla\phi^\dagger)\cdot(\nabla\phi)+\phi^\dagger\left(U(\bm r)+2|\Psi_0|^2-1-\Psi_0^2\tau_+-(\Psi_0^\ast)^2\tau_-\right)\phi-E\tau_3\right].\label{lagrangian}
\end{eqnarray}
Here, $\Omega$ is the volume of the system, and $\tau_\pm$ are given by
\begin{equation}
\tau_+=
\left(
\begin{matrix}
0 & 1\\
0 & 0
\end{matrix}
\right), \ \  
\tau_-=
\left(
\begin{matrix}
0 & 0 \\
1 & 0
\end{matrix}
\right).
\end{equation}

We introduce $N$ spatial positions $\bm r_i$ ($i=1,2,\dots,N$) in
the system, which are referred to as nodes in the literature of the finite
element method \cite{FEM}. We then assign the interpolation function $N_i(\bm r)$
at each $\bm r_i$, which equals unity at $\bm r=\bm r_i$ and linearly
decreases to zero at adjacent nodes of $\bm r_i$.
Namely, the interpolation function satisfies
\begin{equation}
N_i(\bm r_j)=
\left\{
\begin{array}{l}
1 \ \ \ \ \ (i=j),\\
0 \ \ \ \ \ (i\neq j).
\end{array}
\right.
\end{equation}
For example, in one-dimensional case, we define the nodes at $x_i$ ($i=1,2,\dots,
N$). The interpolation function $N_i(x)$ is given by
\begin{equation}
N_i(x)=
\begin{cases}
\frac{x-x_{i-1}}{x_i-x_{i-1}}, & (x_{i-1}\leq x\leq x_i),\\
-\frac{x-x_{i+1}}{x_{i+1}-x_i}, & (x_i\leq x\leq x_{i+1}),\\
0, & (x< x_i, x> x_{i+1} ).
\end{cases}
\end{equation}

Using $N_i(\bm r)$, one can approximately write $u(\bm r)$ and $v(\bm r)$ in the forms
\begin{eqnarray}
u(\bm r)&=&\sum_i u_iN_i(\bm r),\label{disu}\\
v(\bm r)&=&\sum_i v_iN_i(\bm r).\label{disv}
\end{eqnarray}
Substituting Eqs.~(\ref{disu}) and (\ref{disv}) into
Eq.~(\ref{lagrangian}), we obtain
\begin{equation}
L=\sum_{i,j}\left[u_i^\ast\left(K_{i,j}-M_{i,j}^-\right)u_j+v_i^\ast\left(K_{i,j}+M_{i,j}^+\right)v_j-u_i^\ast
	     P_{i,j}v_j-v_i^\ast P_{i,j}^\ast u_j\right],
\label{disL}
\end{equation}
where
\begin{eqnarray}
K_{i,j}&=&\int_{\Omega_{i,j}}d\bm r\ \left(\nabla
				   N_i\right)\cdot\left(\nabla
				   N_j\right),\label{Kij}\\
M_{i,j}^\pm&=&\int_{\Omega_{i,j}}d\bm r\ N_i\left[E\pm(U(\bm
					     r)+2|\Psi_0|^2-1)\right]N_j\nonumber\\
&=&(E\mp 1)\int_{\Omega_{i,j}}d\bm r\ N_iN_j\pm \sum_l
U_l\int_{\Omega_{i,j,l}}d\bm r\ N_iN_jN_l\nonumber\\
&&\pm 2\sum_{l,l'}(f_l-ig_l)(f_{l'}+ig_{l'})\int_{\Omega_{i,j,l,l'}}d\bm
r\ N_iN_jN_lN_{l'},
\label{Mij}\\
P_{i,j}&=&\int_{\Omega_{i,j}}d\bm r\ N_i\Psi_0^2 N_j\nonumber\\
&=&\sum_{l,l'} (f_l+ig_l)(f_{l'}+ig_{l'})\int_{\Omega_{i,j,l,l'}}d\bm r\ N_iN_jN_lN_{l'}.
\label{Pij}
\end{eqnarray}
In obtaining Eqs.~(\ref{Mij}) and (\ref{Pij}), we have expanded $U(\bm r)$ and
$\Psi_0(\bm r)$ as
\begin{eqnarray}
U(\bm r)&=&\sum_l U_l N_l(\bm r),\\
\Psi_0(\bm r)&=&\sum_l (f_l+ig_l) N_l(\bm r).
\end{eqnarray}
Here, $U_l=U(\bm r_l)$, $f_l={\rm Re}\left[\Psi_0(\bm r_l)\right]$, and $g_l={\rm Im}\left[\Psi_0(\bm r_l)\right]$.
In Eqs.~(\ref{Kij})-(\ref{Pij}), $\Omega_{i,j}$, $\Omega_{i,j,l}$, and $\Omega_{i,j,l,l'}$ mean
that the integrations are carried out in the regions where $N_iN_j$,
$N_iN_jN_l$, and $N_iN_jN_lN_{l'}$ are finite, respectively.
The integrations in Eqs.~(\ref{Kij})-(\ref{Pij}) can be evaluated in the
standard manner of the finite element method \cite{FEM}.

Equation~(\ref{disL}) can be rewritten in the matrix form as
\begin{equation}
L=\bm u^\dagger\left(\hat K-\hat M^-\right)\bm u+\bm
 v^\dagger\left(\hat K+\hat M^+\right)\bm v -\bm u\hat P\bm v -\bm
 v^\dagger\hat P^\ast\bm u,
\label{disLm}
\end{equation}
where $\{\bm u\}_i=u_i$, $\{\bm v\}_i=v_i$, $\{\hat K\}_{i,j}=K_{i,j}$,
$\{\hat M^\pm\}_{i,j}=M^\pm_{i,j}$, and $\{\hat P\}_{i,j} = P_{i,j}$.
The equations for $\bm u$ and $\bm v$ are, respectively, obtained from $\delta L/\delta \bm u^\dagger=0$ and $\delta L/\delta \bm
v^\dagger=0$, which give
\begin{eqnarray}
\left(\hat K-E\hat M^-\right)\bm u-\hat P\bm v=0,\label{disBu}\\
\left(\hat K+E\hat M^+\right)\bm v-\hat P\bm u=0.\label{disBv}
\end{eqnarray}

The advantage of using the finite element method is that one can obtain
the solutions by simply diagonalizing Eqs.~(\ref{disBu}) and
(\ref{disBv}) under an appropriate boundary condition, instead of
solving the differential Eqs.~(\ref{Bogoliubovu}) and (\ref{Bogoliubovv}). In the following
sections, we will numerically solve Eqs.~(\ref{disBu}) and
(\ref{disBv}) for given barrier potentials.

\section{tunneling through the rectangular potential barrier}\label{sec3}

In this section, we consider the one-dimensional tunneling problem of Bogoliubov excitations through a
rectangular barrier potential shown in Fig.~\ref{system}. The potential
barrier is given by
\begin{equation}
U(x)=U_0\ \theta\left(\frac{d}{2}-|x|\right),
\label{potential}
\end{equation}
where $\theta(x)$ is the step function. $U_0$ and $d$ describe the
height and width of the
barrier, respectively, and we consider the case of repulsive potential barrier
($U_0>0$). 
In this section, we treat the case when the condensate densities are the
same on both sides of the barrier, as shown in Fig.~\ref{system}. Although this case has been
examined in our previous paper \cite{Tsuchiya}, we give further analyses
for the tunneling of Bogoliubov phonon here.
In Secs.~\ref{sec4} and \ref{sec5}, we will also compare the results in
this section with the case when the condensate density on the left of the
barrier is different from that on the right of the barrier.

In the present case, the GP equation can be solved analytically
\cite{Kagan}, as
\begin{equation}
\bar\Psi_0(\bar x) = 
\begin{cases}
\tanh\left[\frac{1}{\sqrt{2}}\left(|\bar x|-\frac{\bar d}{2}\right)+{\rm
arctanh}\gamma\right], & (|x|\geq d/2),\\
\frac{\beta}{{\rm cn}\left(\sqrt{\frac{K^2+\beta^2}{2}}\bar x,q\right)},
 & (|x|<d/2).
\end{cases}
\label{GPsolution}
\end{equation}
Here, $\bar d\equiv d/\xi$, $K\equiv\sqrt{\beta^2+2(\bar U_0-1)}$, and
$q\equiv K/\sqrt{K^2+\beta^2}$, where $\bar U_0\equiv U_0/\mu$. (${\rm
cn}(x,q)$ is the Jacobi's elliptic function  \cite{Gradshteyn}.)
$\gamma\equiv\bar\Psi_0(\bar x=\bar d/2)$ and $\beta\equiv\bar\Psi_0(0)$
are determined from the boundary conditions in terms of $\Psi_0(x)$ and
$d\Psi_0/dx$ at $x=\pm d/2$, which give
\begin{eqnarray}
\gamma&=&\frac{\beta}{{\rm
 cn}\left(\sqrt{\frac{K^2+\beta^2}{2}}\frac{\bar d}{2},q\right)},\label{bcondition1}\\
\gamma^2&=&\frac{1}{2\bar U_0}\left(\beta^4+2(\bar U_0-1)\beta^2+1\right).
\label{bcondition2}
\end{eqnarray}
The values $\beta$ and $\gamma$ are determined by numerically solving Eqs.~(\ref{bcondition1})
and (\ref{bcondition2}).

To solve Bogoliubov equations (\ref{disBu}) and (\ref{disBv}), we
need asymptotic solutions for $x=\pm\infty$. 
In our tunneling problem, each eigenstate with index $j$ in
Eqs.~(\ref{Bogoliubovu}) and (\ref{Bogoliubovv}) corresponds to
Bogoliubov excitation with energy $E$ injected from one side of the
barrier. In Secs.~\ref{sec3}-\ref{sec5}, we omit the index for
eigenstates for simplicity.
Far from the barrier ($|x|\gg \xi$), the Bogoliubov mode is described by the
plane-wave $(u(x),v(x))=(u_E,v_E)e^{ipx}$.
Substituting this into Eqs.~(\ref{Bogoliubovu}) and (\ref{Bogoliubovv}), one obtains the well-known Bogoliubov
excitation spectrum as \cite{Bogoliubov}
\begin{equation}
E_p=\sqrt{\varepsilon_p(\varepsilon_p+2gn_0)},
\label{Bogoliubovspectrum}
\end{equation}
where $\varepsilon_p=p^2/2m$.
Namely, for a given mode energy $E$, there are four particular solutions
in terms of the momentum $p$, given by
\begin{equation}
p=
\begin{cases}
\pm\sqrt{2m}\sqrt{\sqrt{E^2+(gn_0)^2}-gn_0}\equiv\pm k,\\
\pm i\sqrt{2m}\sqrt{\sqrt{E^2+(gn_0)^2}+gn_0}\equiv\pm i\kappa.
\end{cases}
\label{wave}
\end{equation}
The first two solutions ($p=\pm k$) describe the ordinary propagating waves in the
$\pm x$-directions. The remaining two imaginary solutions ($p=\pm
i\kappa$) describe localized states. We note that while the latter
localized solutions are actually not necessary in a homogeneous system,
we cannot ignore them in the present inhomogeneous system.
The amplitudes of the propagating components are given by
\begin{eqnarray}
\left(
\begin{array}{l}
u^{\rm P}_E\\
v^{\rm P}_E
\end{array}
\right)
=
\left(
\begin{array}{l}
\sqrt{\frac{1}{2L}\left(\frac{\sqrt{E^2+(gn_0)^2}}{E}+1\right)}\\
\sqrt{\frac{1}{2L}\left(\frac{\sqrt{E^2+(gn_0)^2}}{E}-1\right)}
\end{array}
\right)
\equiv\left(
\begin{array}{l}
a\\
b
\end{array}
\right),
\label{ab}
\end{eqnarray}
where $L$ is the system size in the $x$ direction.
On the other hand, the amplitudes for the localized states are given by
$(u_E^{\rm L},v_E^{\rm L})=(-b,a)$.
Thus, in contrast to the propagating solution in Eq.~(\ref{ab}), the
normalization of the localized components becomes negative as
$(u_E^{\rm L})^2-(v^{\rm L}_E)^2=-1/L$. 

Using the propagating solution $(u_E^{\rm P},v_E^{\rm P})e^{\pm ikx}$ and localized
one $(u_E^{\rm L}, v_E^{\rm L})e^{\pm\kappa x}$, we construct the asymptotic forms of
the Bogoliubov wave function for $x\to\pm\infty$.
Assuming that the Bogoliubov phonon is injected from $x=-\infty$, we
obtain the asymptotic solutions as
\begin{equation}
\begin{cases}
\left(\begin{array}{l}
u\\
v
\end{array}\right)=
\left(\begin{array}{l}
a\\
b
\end{array}\right)e^{ikx}+
r\left(
\begin{array}{l}
a\\
b
\end{array}\right)e^{-ikx}+
A\left(\begin{array}{l}
-b\\
a
\end{array}
\right)e^{\kappa x}, & (x\to-\infty)\ ,
\\
\left(\begin{array}{l}
u\\
v
\end{array}\right)=
t\left(
\begin{array}{l}
a\\
b
\end{array}
\right)e^{ikx}+
B\left(
\begin{array}{l}
-b\\
a\end{array}
\right)e^{-\kappa x}, & (x\to\infty)\ .
\end{cases}
\label{asympt}
\end{equation}
Here, $r$ and $t$ are, respectively, the reflection and transmission
amplitudes, which satisfy 
\begin{equation}
|r|^2+|t|^2=1. 
\label{pcons}
\end{equation}
As will be discussed later, this
condition is deeply related to the conservation of energy flux. 
In Eq.~(\ref{asympt}), $A$ and $B$ represent the amplitudes of the localized
components near the potential barrier.

We numerically solve the Bogoliubov coupled Eqs.~(\ref{disBu}) and
(\ref{disBv}) for a given incident energy $E$. In this procedure, the
condensate wave function in Eq.~(\ref{GPsolution}) is used, and the solution
is determined so as to satisfy the asymptotic solution in Eq.~(\ref{asympt}).

\begin{figure}
\centerline{\includegraphics[width=15cm]{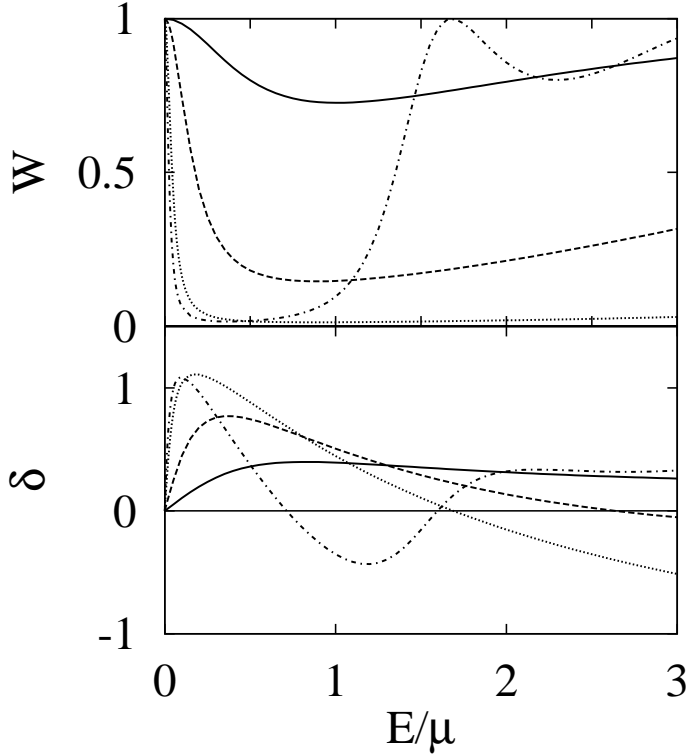}}
\caption{Calculated transmission probability $W$ and phase shift $\delta$ as
 functions of the incident energy $E$ for a
 rectangular potential barrier. We set the width $d$ and height $U_0$ of
 the barrier as $(d,U_0)=(\xi,2\mu)$ (solid line), $(\xi,5\mu)$
 (dashed line), $(\xi,10\mu)$ (dotted line), and $(4\xi,2\mu)$
 (dash-dotted line).}
\label{Wdelta}
\end{figure}

Figure~\ref{Wdelta} shows the calculated transmission probability
$W\equiv|t|^2$, as well as phase shift $\delta\equiv{\rm arg}(t)$, as
functions of the incident energy $E$ for various barrier heights and widths.
We call attention to the characteristic features of $W$ and $\delta$ in
the low-energy region ($E/\mu \lesssim 0.5$). One can clearly see the anomalous tunneling behavior 
discussed in \cite{Kovrizhin1,Kovrizhin2,Kagan} in
Fig.~\ref{Wdelta}. Namely, below a certain incident energy ($E/\mu\sim 0.5$),
$W$ increases and $\delta$ decreases with decreasing $E$, in contrast to
the behaviors above that energy ($W$ decreases and $\delta$ increases as
$E$ decreases).
Furthermore, $W$ and $\delta$ approach unity and zero in the low-energy limit $E\to
0$, respectively, irrespective of the values of $d$ and $U_0$.
When the incident energy $E$ is very large ($E\gg\mu$), since the
Bogoliubov phonon loses its collective nature, the tunneling property becomes
close to that of a single particle.

We note that the {\it perfect transmission} of Bogoliubov
phonon ($W\to 1,\delta\to 0$) shown in the low-energy limit in Fig.~\ref{Wdelta} is
quite different from the typical tunneling
properties of a single particle, where $W$ and $\delta$ approach
0 and $-\pi/2$ in the low-energy limit, respectively. Namely, in the
latter case the particle is completely reflected by the potential barrier
\cite{LandauLifshitz}. 

We also note that the energy region in which $W$ and $\delta$
exhibit the anomalous tunneling behavior ($W$ increases and $\delta$
decreases with decreasing $E$)
depends on the height $U_0$ and width $d$ of the potential 
barrier. This region becomes narrower for higher and wider potential
barrier, as shown in Fig.~\ref{Wdelta}.

\begin{figure}
\centerline{\includegraphics[width=15cm]{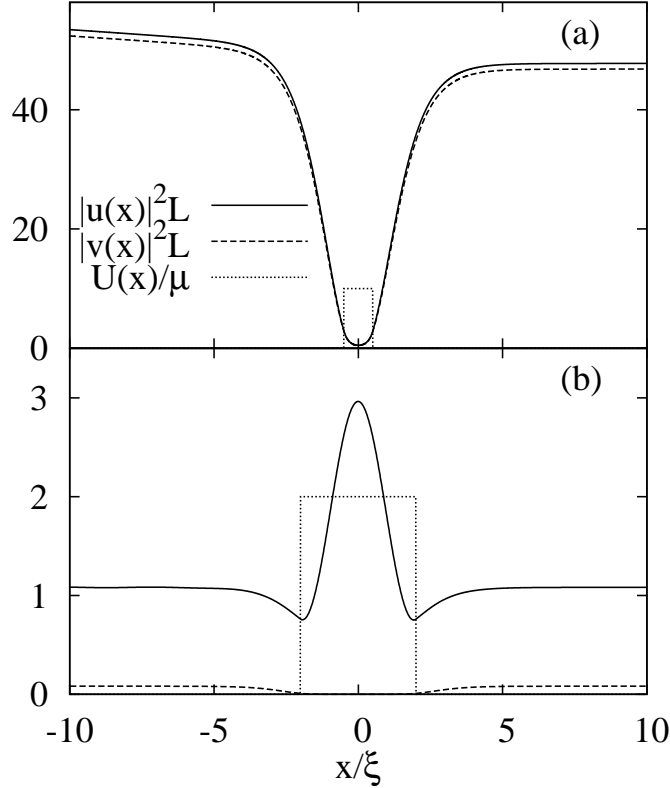}}
\caption{Spatial variation in the Bogoliubov wave function
 $(u(x),v(x))$. (a) $E/\mu=0.01\ll 1$ (anomalous tunneling). We set
 $(d,U_0)=(\xi,10\mu)$.
(b) $E/\mu=1.68$. We set $(d,U_0)=(4\xi,2\mu)$. In this case, the
 resonance tunneling ($W=1$) is realized, as shown in
 Fig.~\ref{Wdelta} (see the dash-dotted line). The dotted line
 is the potential barrier $U(x)$ in units of $\mu$.}
\label{wf}
\end{figure}

In Fig.~\ref{Wdelta}, we find that $W=1$ is
also obtained at finite energy $(E/\mu=1.68)$ in the case of
$(d,U_0)=(4\xi,2\mu)$, due to the resonance tunneling effect. To see the
difference between the resonance tunneling
effect and the anomalous
tunneling effect, we show in Fig.~\ref{wf} the wave functions in the two cases.
In the case of resonance tunneling, one sees that while $|u|^2$ is
enhanced in the barrier, $|v|^2$ is suppressed there. The peak of $|u|^2$ is a
clear signature of the formation of a resonance state.
The suppression of $|v|^2$ indicates that the Bogoliubov excitation
behaves like a single particle during the tunneling through the barrier.
In contrast, in the case of the anomalous tunneling, both $|u|^2$ and
$|v|^2$ simply become small in the barrier and almost
coincide with each other. Indeed, it was shown in Refs.~\cite{Kato,Ohashi} that
$u(x)$ and $v(x)$ reduce to the condensate wave function $\Psi_0(x)$ in
the low-energy limit.  
The difference mentioned above indicates that the anomalous tunneling and resonance tunneling
are different phenomena.

We briefly note that, as shown in Ref.~\cite{Ohashi}, the anomalous
tunneling effect originates from the fact that the wave functions of a
Bogoliubov phonon with a small momentum $p$ has the same form as the
condensate wave function in the {\it supercurrent state}, accompanied by
a finite superflow $J_s=n_0p/m$. Recently, Morgan {\it et al.} \cite{Morgan}
have presented a modified Bogoliubov theory where the wave function of
Bogoliubov mode is constructed so as to be orthogonal to the solution
obtained from the GP equation. Since their formalism does not affect the
current-carrying component of the Bogoliubov wave function (which
dominates the anomalous tunneling phenomenon), the perfect transmission
of low-energy Bogoliubov phonon is still expected to occur.  Thus, the
anomalous tunneling phenomenon does not depend on the definition of the
wave function of Bogoliubov mode.

Propagation of Bogoliubov phonon is accompanied by quasiparticle current
$J_{\rm q}$, as well as energy flux $Q_{\rm q}$. When one uses the
asymptotic solutions in Eq.~(\ref{asympt}), they are given by
\begin{eqnarray}
J_{\rm q}=
\left\{
\begin{array}{l}
\frac{k}{mL}(1-|r|^2)-\frac{2ab}{mL}e^{\kappa x}\left(\kappa\ {\rm
						 Im}\left[A(e^{-ikx}+r^\ast
						     e^{ikx})\right]+k\ {\rm Re}\left[A(e^{-ikx}-r^\ast e^{ikx})\right]\right),\\
\quad\quad\quad\quad(x\ll-\xi),\\
\frac{k}{mL}|t|^2-\frac{2ab}{mL}e^{-\kappa x}\left(\kappa\ {\rm
					     Im}\left[tB^\ast
						 e^{ikx}\right]+k\ {\rm Re}
					     \left[tB^\ast
					      e^{ikx}\right]\right),\quad\quad(x\gg\xi),
\end{array}
\right.
\label{asymptJq}
\end{eqnarray}
\begin{equation}
Q_{\rm q}=
\begin{cases}
\frac{kE}{m}(a^2+b^2)(1-|r|^2),\quad\quad(x\ll -\xi),\\
\frac{kE}{m}(a^2+b^2)|t|^2,\quad\quad\quad\quad\quad(x\gg\xi).
\end{cases}
\label{asymptQ}
\end{equation}
The detailed definitions of $J_{\rm q}$ and $Q_{\rm q}$ are summarized
in Appendix B.
Since the energy flux $Q_{\rm q}$ is conserved (see Appendix B), one
obtains $|r|^2+|t|^2=1$ from Eq.~(\ref{asymptQ}). Using this, we find
that the quasiparticle current $J_{\rm q}$ is also conserved in both
limits $x=\pm\infty$ as $J_{\rm q}(x=-\infty)=k(1-|r|^2)/mL=J_{\rm
q}(x=\infty)=k|t|^2/mL$. However, except for the limits $x=\pm\infty$,
the last terms in Eq.~(\ref{asymptJq}) become finite, which come from
the coupling between the propagating and localized components in
Eq.~(\ref{asympt}). As a result, while $Q_{\rm q}$ is conserved
everywhere, we expect that $J_{\rm q}$ is not conserved near the barrier.

\begin{figure}
\centerline{\includegraphics[width=15cm]{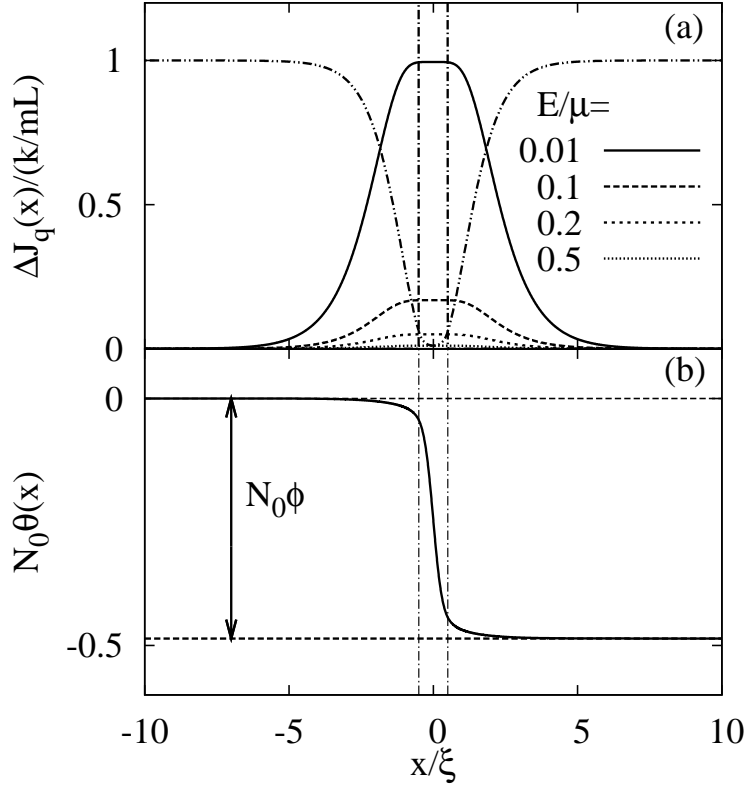}}
\caption{(a): Excess quasiparticle current $\Delta J_{\rm q}(x)\equiv
 J_{\rm q}(x)-J_{\rm q}(x=-\infty)$ when $(d,U_0)=(\xi,10\mu)$. The
 dash-double dotted line is the condensate density $n_{\rm s}(x)$ in
 units of $n_0$. (b): Phase $\theta(x)$ of the condensate wave function $\Psi_0(x)$
 created by the tunneling of Bogoliubov phonon. We set
 $\theta(x=-\infty)=0$. $\phi$ is the phase difference between
 condensates at $x=\pm\infty$. $N_0=n_0L$ is the number of condensate
 atoms. In both the panels, the dash-dotted line indicates the region of
 the potential barrier.
}
\label{Jqphase}
\end{figure}

To see the non-conserving behavior of $J_{\rm q}$, we directly evaluate
it using the solution of Bogoliubov equations (\ref{disBu}) and
(\ref{disBv}). As shown in Fig.~\ref{Jqphase}(a), we obtain the {\it
excess} quasiparticle current 
\begin{equation}
\Delta J_{\rm q}(x)\equiv J_{\rm q}(x)-J_{\rm q}(x=-\infty)
\end{equation}
near the barrier. Namely, when the Bogoliubov phonon approaches the
barrier, $J_{\rm q}$ is enhanced. $J_{\rm q}$ is constant in the
barrier, and it decreases to be $J_{\rm q}(\infty)=J_{\rm q}(-\infty)$ when the phonon
goes away from the barrier. In Fig.~\ref{Jqphase}(a), the enhancement of $\Delta J_{\rm q}$ occurs
near the barrier where the condensate density $n_{\rm s}(x)$ deviates
from $n_0[=n_{\rm s}(x=\pm\infty)]$.

\begin{figure}
\centerline{\includegraphics{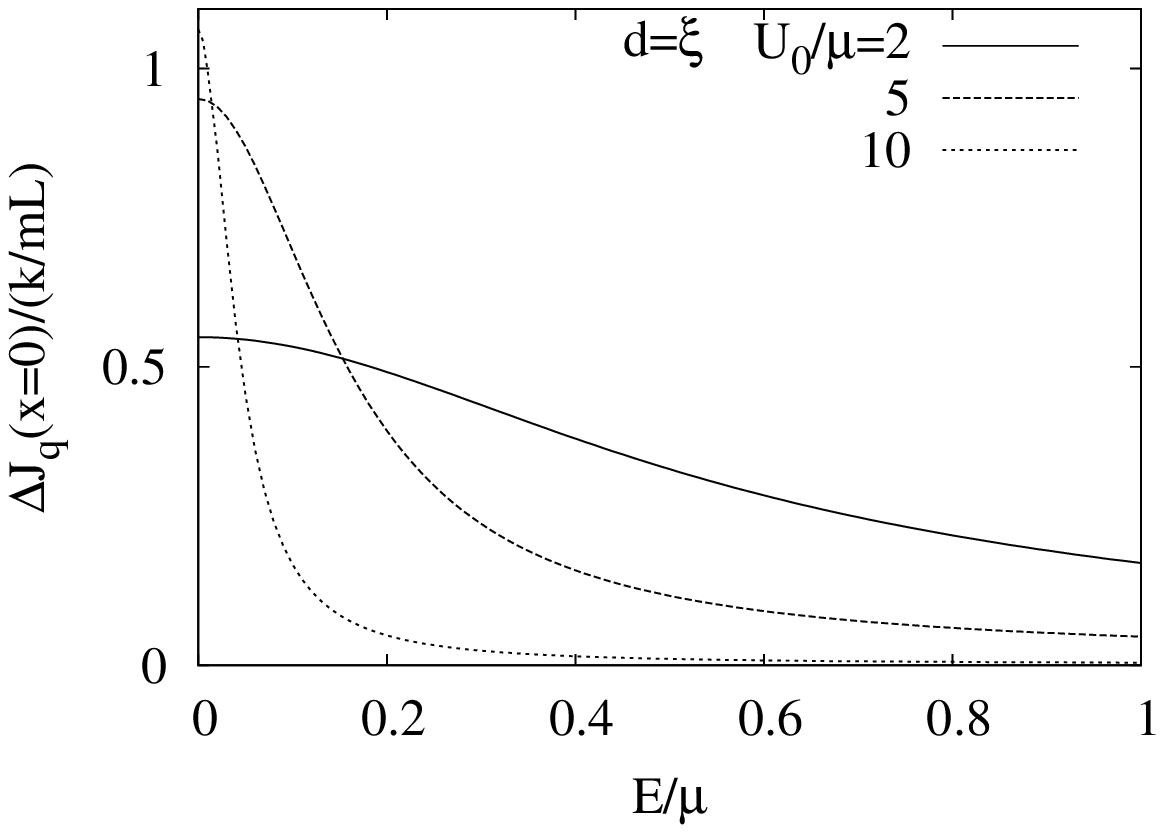}}
\caption{Excess quasiparticle current in the potential barrier $\Delta
 J_{\rm q}(x=0)$ as a function of the incident energy $E$.}
\label{dJqx0}
\end{figure}

In Fig.~\ref{Jqphase}(a), we find that the enhancement of $\Delta J_{\rm q}$ is
more pronounced for lower incident energy $E$. In addition, as shown in
Fig.~\ref{dJqx0}, the excess quasiparticle current is more remarkable
when the barrier is high, although the energy region where $\Delta J_{\rm
q}(x=0)$ is large is narrower for larger $U_0$. 
Since $\Delta J_{\rm q}(x=0)/(k/mL)$ approaches a constant value, we also find from Fig.~\ref{dJqx0}
that $\Delta J_{\rm q}(x=0)$ is proportional to the incident momentum $k$ in the low-energy limit.
[Note that $\Delta J_{\rm q}(x=0)$ in Fig.~\ref{dJqx0} is normalized by the incident quasiparticle current
$k/mL$.]

The enhancement of quasiparticle current near the potential barrier implies
that more quasiparticles than those carried in the incident current impinge on the
barrier. Apparently, this is expected to lead to the increase in the transmission
probability of quasiparticles.
Indeed, comparing the result for $(d,U_0)=(\xi,10\mu)$ in
Fig.~\ref{Jqphase} with the corresponding result in Fig.~\ref{Wdelta},
one finds that the energy region where the anomalous enhancement of transmission probability
is obtained ($E\lesssim 0.1\mu$) coincides with the region where the excess
quasiparticle current $\Delta J_{\rm q}(x=0)$ is remarkable.

As shown in Ref.~\cite{Tsuchiya}, the excess quasiparticle current is
supplied from the condensate. Namely, the transmission of Bogoliubov
phonon is considered to be assisted by the supply of excess current from the
condensate. Thus, in a sense, the mechanism of the anomalous tunneling may be considered as
a kind of screening effect by Bose condensate.
This argument partially explains the physical mechanism of the anomalous tunneling effect discussed in
Refs.~\cite{Kovrizhin1,Kovrizhin2,Kagan}. However, apart from the
enhancement of low-energy transmission probability, this argument is not
enough to explain the {\it perfect} transmission in the low-energy
limit. In this regard, in Ref.~\cite{Ohashi}, we have shown
that the perfect transmission can be understood as a result of the
supercurrent behavior of low-energy Bogoliubov phonon.

In Ref.~\cite{Tsuchiya}, it was found that the counterflow of supercurrent is
induced near the potential barrier due to a back-reaction effect of
quasiparticle current, which restores the conservation of total current.
The induction of supercurrent indicates that the phase of the BEC order
parameter $\Psi_0(x)$ is twisted by quasiparticle current as
\begin{equation}
\Psi_0(x)\to e^{i\theta(x)}\Psi_0(x).
\label{thetaansatz}
\end{equation}
(Here, we assume that the amplitude of the condensate wave function is unchanged.)
The induced supercurrent by this phase modulation is given by
\begin{equation}
\Delta J_{\rm s}(x)=\frac{n_{\rm s}(x)}{m}\partial_x\theta(x).
\label{deltaJs}
\end{equation}
As shown in Appendix B, $\Delta J_{\rm s}(x)$ is related to the excess
quasiparticle current $\Delta J_{\rm q}$ as 
\begin{equation}
\Delta J_{\rm s}(x)=-\Delta J_{\rm q}(x).
\label{deltaJs2}
\end{equation}
[Here, we set $\langle\alpha_j^\dagger\alpha_j\rangle=1$ in Eq.~(\ref{contJs}) assuming that one
Bogoliubov excitation is injected.]
As a result, the phase $\theta(x)$ is evaluated to be
\begin{equation}
\theta(x)=-m\int_{-\infty}^x dx^{\prime} \frac{\Delta J_{\rm q}(x^{\prime})}{n_{\rm
		   s}(x^{\prime})}\ .\label{theta}
\end{equation}
Namely, the phase modulation is caused by the excess quasiparticle
current $\Delta J_{\rm q}$.
The assumption in Eq.~(\ref{thetaansatz}) is valid as long as
$\theta(x)$ is small, because the change in the amplitude of the
condensate wave function gives higher-order corrections.
Since $\theta(x)$ is inversely proportional to the number of condensate atoms
$N_0$, as shown below, $\theta(x)$ is negligibly small, so that the
assumption in Eq.~(\ref{thetaansatz}) is justified.
As discussed in Appendix \ref{B}, the inclusion of the back-reaction effect
of quasiparticles on condensates requires the modification of the GP
equation as Eq.~(\ref{gGP}). In the present case, the new condensate wave
function including the back-reaction effect is perturbatively obtained with use of
the ansatz in Eq.~(\ref{thetaansatz}) without solving Eq.~(\ref{gGP}).

Figure~\ref{Jqphase}(b) shows $\theta(x)$ when the Bogoliubov phonon
is injected from $x=-\infty$. The spatial variation in
the phase $\theta(x)$ is remarkable near and in the barrier, where large
excess current $\Delta J_{\rm q}$ is obtained [see Fig.~\ref{Jqphase}(a)].

\begin{figure}
\centerline{\includegraphics{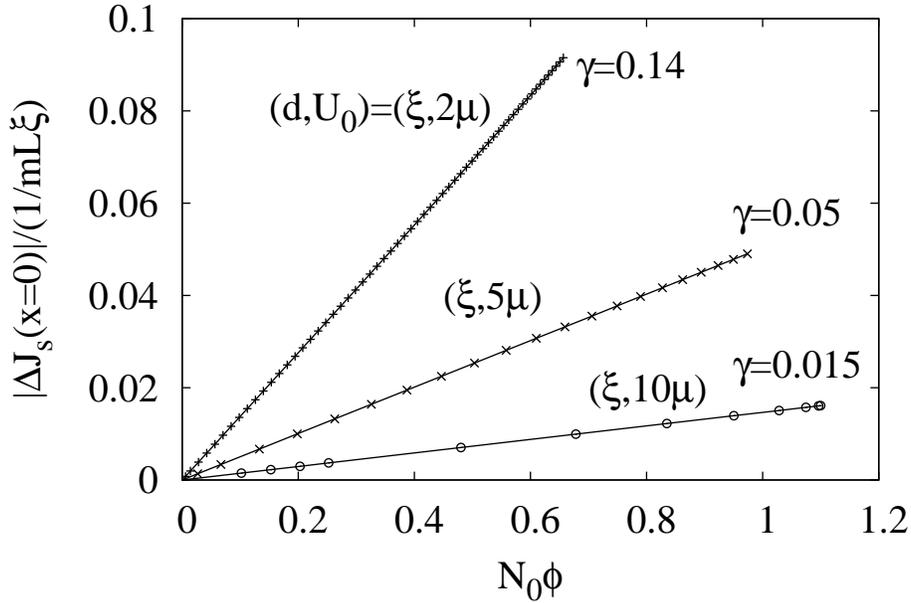}}
\caption{Induced supercurrent $\Delta J_{\rm s}(x=0)$ as a function
 of the relative phase $\phi$ across the potential barrier. The slopes of the lines are
 $\gamma=0.14$, $0.050$, and $0.015$ for $(d,U_0)=(\xi,2\mu)$,
 $(\xi,5\mu)$, and $(\xi,10\mu)$, respectively.}
\label{josephson}
\end{figure}

Figure~\ref{josephson} shows the magnitude of induced supercurrent at
$x=0$ as a function of $\phi\equiv\theta(-\infty)-\theta(x)$ [see
Fig.~\ref{Jqphase}(b)].
We clearly see that $|\Delta J_{\rm s}(x=0)|$ satisfies the ordinary Josephson current relation \cite{Josephson}
\begin{equation}
I(\phi)=I_J\sin\phi\simeq I_J\phi,\ \ \ \ (\phi\ll 1).
\end{equation}
(In our case, since $\phi$ is proportional to the inverse of the total
number of Bose-condensed particles $N_0$, so that $\phi\ll 1$.)
The Josephson critical current $I_J$ in the present case has the form,
\begin{equation} 
I_J=\gamma\left(\frac{n_0}{m\xi}\right),
\end{equation}
where $\gamma$ is determined from the slope of the lines in Fig.~\ref{josephson}.
This result means that the Josephson critical current $I_J$ may be
evaluated from the analysis of quasiparticle tunneling without directly
examining the Josephson current.

Finally, we remark that the tunneling properties of Bogoliubov excitation
discussed in this section suggest an important role of Bogoliubov phonons on the fluctuation
of the relative phase between two condensates at finite temperatures.
When Bogoliubov phonons are excited on both sides of the barrier at finite
temperatures, they tunnel through the potential barrier and twist the relative
phase. This is expected to lead the fluctuation of the phase difference
between the condensates on the left and right of the barrier. 
In particular, large phase fluctuations may be induced in the
temperature region where the population of Bogoliubov phonon becomes dominant.
This phase fluctuation due to the tunneling of Bogoliubov phonons could
be observed in a BEC in a double-well potential, where the thermally induced fluctuations of
the relative phase between two condensates were recently observed \cite{Gati}.

\section{tunneling between condensates with different condensate densities}
\label{sec4}

\begin{figure}
\centerline{\includegraphics{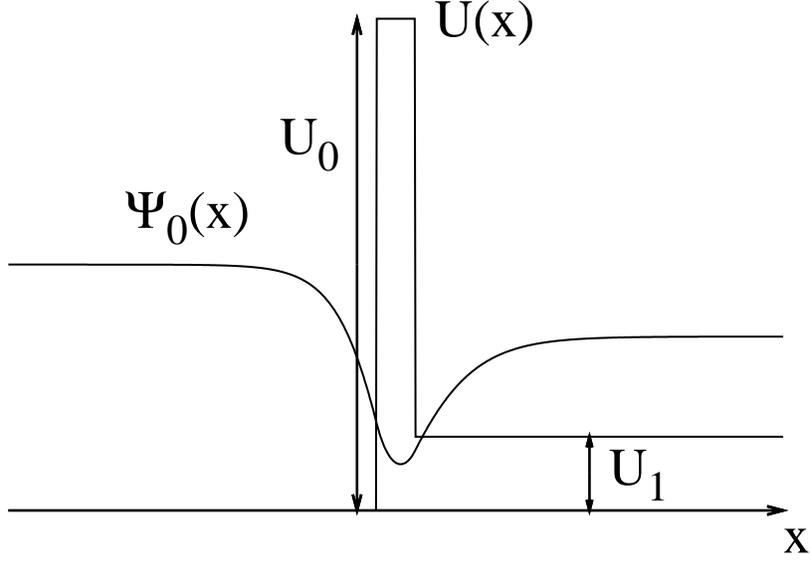}}
\caption{Schematic of the system in the presence of the barrier
 potential in Eq.~(\ref{steppotential}).}
\label{stepsystem}
\end{figure}

In Sec.~\ref{sec3}, we considered tunneling properties of Bogoliubov
phonons through the rectangular potential barrier in the case when the
left and right of the barrier have the same condensate densities.
In this section, we consider the more general case when the condensate
densities are different between the right and left of the barrier.
This situation is achieved by simply imposing a uniform potential on the
right side of the barrier, as 
\begin{equation}
U(x)=U_0\theta\left(\frac{d}{2}-|x|\right)+U_1\theta\left(x-\frac{d}{2}\right).
\label{steppotential}
\end{equation}
In this case, the condensate density at $x\to\infty$ is given by
\begin{equation}
\tilde n_0=\Psi_0(x=\infty)^2=
\begin{cases}
\frac{1}{g}(\mu- U_1), &  (0\leq U_1<\mu),\\
0, & (U_1\geq\mu).
\end{cases}
\end{equation}
In this section, we consider the case of $0\leq U_1<\mu$. The case of
$U_1\geq\mu$ will be discussed in Sec.~\ref{sec5}. The barrier potential,
as well as the condensate wave function $\Psi_0(x)$, is schematically shown in
Fig.~\ref{stepsystem}.

To solve Bogoliubov equations (\ref{disBu}) and (\ref{disBv}), we
construct the condensate wave function $\Psi_0(x)$, as well as the
asymptotic solutions at $x=\pm\infty$. The former is analytically
obtained from Eq.~(\ref{GP}) as
\begin{eqnarray}
\bar\Psi_0(\bar x)=
\begin{cases}
\tanh\left[-\frac{1}{\sqrt{2}}\left({\bar x}+\frac{\bar d}{2}\right)+{\rm
      arctanh}\gamma_L\right], & (x<-\frac{d}{2}),\\
\frac{\beta}{{\rm
 cn}\left(\sqrt{\frac{K^2+\beta^2}{2}}(\bar x-\bar x_0),q\right)}, & (|x|\le d/2)\\
\sqrt{1-\bar U_1}\tanh\left[\sqrt{\frac{1-\bar
		       U_1}{2}}\left(\bar x-\frac{\bar d}{2}\right)+{\rm
      arctanh}\left(\frac{\gamma_R}{\sqrt{1-\bar U_1}}\right)\right], & (x>\frac{d}{2}),
\end{cases}
\label{Psi0step}
\end{eqnarray}
where $\bar U_1\equiv U_1/\mu$, $\gamma_L\equiv\bar\Psi_0(-\bar d/2)$,
and $\gamma_R\equiv\bar\Psi_0(\bar d/2)$. $x_0$ satisfies the conditions
$\bar\Psi(\bar x_0)=\beta$ and $d\Psi_0(x)/dx|_{x=x_0}=0$. 
$x_0$, $\beta$, $\gamma_L$, and $\gamma_R$ are determined from the equations,
\begin{eqnarray}
\gamma_R&=&\frac{\beta}{{\rm
 cn}\left(\sqrt{\frac{K^2+\beta^2}{2}}(\frac{\bar d}{2}-\bar
     x_0),q\right)},\label{bd1}\\
\gamma_L&=&\frac{\beta}{{\rm
 cn}\left(\sqrt{\frac{K^2+\beta^2}{2}}(\frac{\bar d}{2}+\bar
     x_0),q\right)},\label{bd2}\\
\gamma_R^2&=&\frac{1}{2(\bar U_0-\bar U_1)}\left(\beta^4+2(\bar
					  U_0-1)\beta^2+(1-\bar
					  U_1)^2\right),\label{bd3}\\
\gamma_L^2&=&\frac{1}{2\bar U_0}\left(\beta^4+2(\bar
					  U_0-1)\beta^2+1\right).\label{bd4}
\end{eqnarray}
Equations~(\ref{bd1})-(\ref{bd4}) are derived from the boundary
conditions at $x=\pm d/2$.

The asymptotic solutions of the Bogoliubov equations at $x=\pm\infty$
are obtained in the same manner as in Sec.~\ref{sec3}. Assuming that the
Bogoliubov phonon with the energy
$E=\sqrt{\varepsilon_p(\varepsilon_p+2g\tilde n)}$ is injected from
$x=-\infty$, we have
\begin{equation}
\begin{cases}
\left(
\begin{array}{l}
u\\
v
\end{array}\right)=
\left(\begin{array}{l}
a\\
b
\end{array}\right)e^{ikx}+
r\left(
\begin{array}{l}
a\\
b
\end{array}\right)e^{-ikx}+
A\left(\begin{array}{l}
-b\\
a
\end{array}
\right)e^{\kappa x}, & (x\to-\infty),\\
\left(\begin{array}{l}
u\\
v
\end{array}\right)=
t\left(
\begin{array}{l}
a_R\\
b_R
\end{array}
\right)e^{ik_Rx}+
B\left(
\begin{array}{l}
-b_R\\
a_R\end{array}
\right)e^{-\kappa_R x}, & (x\to\infty).
\end{cases}
\label{asymptstep}
\end{equation}
Here, $k$, $\kappa$, and $(a,b)$ are given in Eqs.~(\ref{wave})
and (\ref{ab}). The parameters appearing in the asymptotic
solution at $x=\infty$ are given by
\begin{eqnarray}
k_R&=&\sqrt{2m}\sqrt{\sqrt{E^2+(g\tilde n_0)^2}-g\tilde n_0},\\
\kappa_R&=&\sqrt{2m}\sqrt{\sqrt{E^2+(g\tilde n_0)^2}+g\tilde n_0},\\
\left(
\begin{array}{l}
a_R\\
b_R
\end{array}
\right)
&=&
\left(
\begin{array}{l}
\sqrt{\frac{1}{2L}\left(\frac{\sqrt{E^2+(g\tilde n_0)^2}}{E}+1\right)}\\
\sqrt{\frac{1}{2L}\left(\frac{\sqrt{E^2+(g\tilde n_0)^2}}{E}-1\right)}
\end{array}
\right).
\end{eqnarray}

Using the condensate wave function in Eq.~(\ref{Psi0step}), we
numerically solve the Bogoliubov Eqs.~(\ref{disBu}) and
(\ref{disBv}) so as to satisfy Eq.~(\ref{asymptstep}). Once the wave
function $(u(x),v(x))$, as well as $r$ and $t$, are determined, we can
calculate the transmission probability from the conserving energy flux
$Q_{\rm q}$. The energy flux $Q_{\rm q}$ at $x\gg \xi$ is
given by
\begin{equation}
Q_{\rm q}=\frac{k_RE}{mL}(a_R^2+b_R^2)|t|^2.
\label{stepQq}
\end{equation}
The transmission (reflection) probability $W$
($R$) is conveniently defined as the ratio of the incident and
transmitted (reflected) components of $Q_{\rm q}$. From Eqs.~(\ref{asymptQ})
and (\ref{stepQq}), we obtain
\begin{eqnarray}
W&=&\frac{k_R(a_R^2+b_R^2)}{k(a^2+b^2)}|t|^2,\label{WQ}\\
R&=&|r|^2.\label{RQ}
\end{eqnarray}
Equations (\ref{WQ}) and (\ref{RQ}) satisfy the relation
$R+W=1$ because of the conservation of $Q_{\rm q}$ as proved in Appendix~\ref{B}.

We note that, when we calculate the transmission probability from the
quasiparticle current, we obtain a different result from Eqs.~(\ref{WQ})
and (\ref{RQ}). Using the expression for the quasiparticle current at
$x\gg \xi$,
\begin{equation}
J_{\rm q}=\frac{k_R}{mL}|t|^2-\frac{2a_Rb_R}{m}e^{-\kappa_R
 x}(\kappa_R{\rm Im}\left[tB^\ast e^{ik_Rx}\right]+k_R{\rm
 Re}\left[tB^\ast e^{ik_Rx}\right]),
\label{stepJq}
\end{equation}
and Eq.~(\ref{asymptJq}), we define the ``transmission (reflection)
probability'' $W_J$ ($R_J$) as the ratio of the incident and transmitted
(reflected) components of $J_{\rm q}(x=\pm\infty)$. Then, we find
\begin{eqnarray}
W_J&=&\frac{k_R}{k}|t|^2=\frac{a^2+b^2}{a_R^2+b_R^2}W,\label{WJ}\\
R_J&=&|r|^2=R.\label{RJ}
\end{eqnarray}
Since $W_J>W$, Eqs.~(\ref{WJ}) and (\ref{RJ}) do {\it not} satisfy the condition
$R_J+W_J=1$, unless $U_1=0$. 
This is because of the fact that $J_{\rm q}$ is not conserved, as
discussed in Sec.~\ref{sec3} and Appendix \ref{B}. 
When $U_1=0$ (this case was discussed in Sec.~\ref{sec3}), the breakdown
of the conservation of $J_{\rm q}$ is restricted to the region near the
barrier. Namely, all the supplied component $\Delta J_{\rm q}$ is completely absorbed
after the quasiparticle is transmitted in the right condensate, as shown in Fig.~\ref{Jqphase}. 
As a result, the transmission probability, which is defined using
$J_{\rm q}(x=\pm\infty)$, is not affected by this non-conserving
character of $J_{\rm q}$. On the other hand, the fact of $W_J+R_J>0$ when $U_1>0$ indicates that the non-conserving behavior
of $J_{\rm q}$ remains even at $x=\infty$.

\begin{figure}
\centerline{\includegraphics[width=15cm]{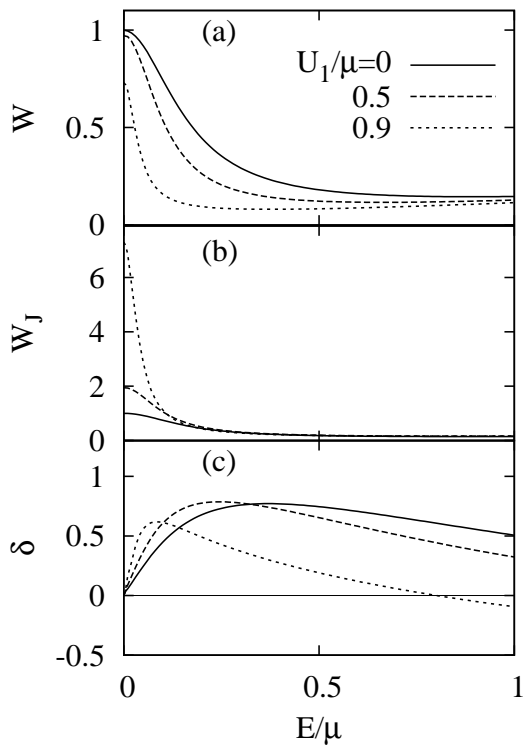}}
\caption{Transmission probability $W$ (a) calculated from the energy flux,
 and $W_J$ (b) calculated from the quasiparticle current
 $J_{\rm q}$, and phase shift $\delta$ (c). We set $(d,U_0)=(\xi,5\mu)$.}
\label{Wdeltastep}
\end{figure}

Figure~\ref{Wdeltastep} shows the calculated transmission probability
$W$, $W_J$ in Eq.~(\ref{WJ}), and the phase shift $\delta\equiv {\rm arg}(t)$.
While the phase shift $\delta$ approaches 0 in the low-energy limit
irrespective of the value of $U_1$, the perfect transmission
($W\to 1$ in the low-energy limit) is absent when $U_1>0$.
In Ref.~\cite{Watabe}, Watabe and Kato have obtained the analytic
expressions for $W$ and $\delta$ in the low-energy limit for arbitrary potential barrier shape.
According to their results, $W$ and $\delta$ read \cite{Watabe}
\begin{equation}
W\to\frac{4\sqrt{1-\bar U_1}}{(1+\sqrt{1-\bar
 U_1})^2},\quad\quad\delta\to 0, \quad\quad(E\to
 0).
\label{Watabe}
\end{equation}
These results can be also obtained in the case of a $\delta$-function
potential barrier \cite{Tsuchiya2}.
Our results in Fig.~\ref{Wdeltastep} are consistent with their
earlier results in Eq.~(\ref{Watabe}).
Equation~(\ref{Watabe}) shows that the transmission probability
$W$ becomes less than unity when $U_1>0$.
As pointed out in Ref.~\cite{Watabe}, the low-energy behaviors of $W$
and $\delta$ are determined only by the potential difference at
$x=\pm\infty$ ($U_1$ in the present case), and
they do not depend on the detail of the potential barrier in the middle.

\begin{figure}
\centerline{\includegraphics[width=15cm]{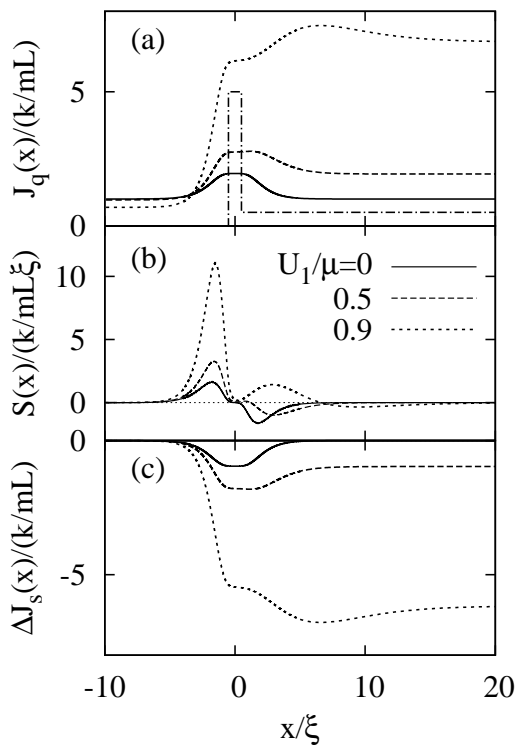}}
\caption{Spatial variation in quasiparticle current $J_{\rm q}(x)$
 (a), source term $S(x)$ (b), and induced
 supercurrent $\Delta J_{\rm s}$ (c) when $E=0.01\mu$. 
 We set $(d,U_0)=(\xi,5\mu)$. The dash-dotted line in (a) shows the
 potential barrier $U(x)$ in units of $\mu$ when $U_1=0.5\mu$.}
\label{JqSJstep}
\end{figure}

In Fig.~\ref{Wdeltastep}(b), one finds that 
$W_J$ is remarkably enhanced to be larger than unity in the low-energy
region. 
To see the relation of this large enhancement and the non-conserving character of $J_{\rm q}$, we show the
spatial variation in $J_{\rm q}$ in Fig.~\ref{JqSJstep}(a). Comparing this
result with Fig.~\ref{Jqphase}, we find that the excess quasiparticle current
$\Delta J_{\rm q}$ remains finite even far away from the barrier ($x\gg
d$) when $U_1>0$. This excess current is found to be supplied from the
condensate through the source term $S(x)$ defined in Eq.~(\ref{source}),
as shown in Fig.~\ref{JqSJstep}(b). Figure \ref{JqSJstep}(b) also shows
that this supply dominantly occurs in front of the barrier ($-5\lesssim
x/\xi\lesssim 0$). (Note that the phonon is injected from
$x=-\infty$.) As a result, $W_J$ given by the ratio of incident and
transmitted quasiparticle current is remarkably enhanced.

As discussed in Sec.~\ref{sec3}, the excess component $\Delta J_{\rm
q}(x)=J_{\rm q}(x)-J_{\rm q}(x=-\infty)$ is cancelled out by the counter
flow of supercurrent to conserve the total current.
As shown in Fig.~\ref{JqSJstep}(c), the induced supercurrent remains
finite even at $x\to\infty$, which is in contrast to the case of $U_1=0$,
where $\Delta J_{\rm s}$ is only finite near the barrier.
The reason for this can be considered as follows: as discussed in
Sec.~\ref{sec3}, Bogoliubov phonons twist the condensate phase when they
tunnel through a potential barrier.
In addition, Bogoliubov phonons can be regarded as quantized
oscillations of the phase of the condensate wave function
\cite{Pitaevskii}. 
Since the phase stiffness is weak on the right side of
the barrier due to the small condensate density, the transmitted
Bogoliubov phonons can easily twist the phase of the right condensate
when $U_1>0$.
This leads to the induction of counter superflow far away from the
barrier. Indeed, $|\Delta J_{\rm s}(x\gg\xi)|$ and
$J_{\rm q}(x\gg\xi)$ are larger for larger $U_1$, as shown in
Fig.~\ref{JqSJstep}.

\section{tunneling between superfluid and normal regions}
\label{sec5}

In this section, we consider the case when the condensate density at
$x\ll -\xi$ is absent. To realize this situation in a simple manner, we
use the potential
\begin{equation}
U(x)=U_2\theta(-x)
\label{NSpotential}
\end{equation}
with $U_2\geq\mu$. In what follows, we call the negative $x$ side the
normal region and the positive $x$ side the superfluid region.
In the normal region, Bogoliubov excitations reduce to free atoms,
having the energy
\begin{equation}
E_p^{\rm s}=\varepsilon_p+(U_2-\mu).
\end{equation}

Here, we discuss two different tunneling problems, i.e., tunneling
of atoms from the normal region to the superfluid region (N-S tunneling), and
the tunneling of Bogoliubov excitations from the superfluid region to the normal
region (S-N tunneling). 
These two cases enable us to study how free atoms are
injected into a condensate and emitted from the surface of the condensate.
We note that these tunneling problems are
analogous to the quantum evaporation and condensation at a free surface in
superfluid $^4$He \cite{Dalfovo}.

We first consider the N-S tunneling. 
The analytic solution of the GP Eq. (\ref{GP}) is given by
\begin{eqnarray}
\bar\Psi_0(\bar x)=
\begin{cases}
-\frac{\lambda}{\sinh\left(\lambda\frac{\bar x}{\sqrt{2}}-C\right)}, & (x<0),\\
\tanh\left(\frac{\bar x}{\sqrt{2}}+{\rm arctanh}\alpha\right), & (x\geq 0),
\end{cases}
\end{eqnarray}
where $\alpha=1/\sqrt{2\bar U_2}$, $\lambda=\sqrt{2(\bar U_2-1)}$, and
$C=\log[(\lambda+\sqrt{\alpha^2+\lambda^2})/\alpha]$.

The asymptotic solution of the Bogoliubov equations has the form
\begin{eqnarray}
\begin{cases}
\left(\begin{array}{l}
u\\
v
\end{array}\right)=
\left(\begin{array}{l}
1\\
0
\end{array}\right)\frac{e^{ik_Lx}}{\sqrt{L}}+
r\left(
\begin{array}{l}
1\\
0
\end{array}\right)\frac{e^{-ik_Lx}}{\sqrt{L}}+
A\left(\begin{array}{l}
0\\
1
\end{array}
\right)\frac{e^{\kappa_L x}}{\sqrt{L}}, & (x\to-\infty),\\
\left(\begin{array}{l}
u\\
v
\end{array}\right)=
t\left(
\begin{array}{l}
a\\
b
\end{array}
\right)e^{ikx}+
B\left(
\begin{array}{l}
-b\\
a\end{array}
\right)e^{-\kappa x}, & (x\to\infty),
\end{cases}
\label{step2asympt}
\end{eqnarray}
where $k$, $\kappa$, and $(a,b)$ are given in Eqs.~(\ref{wave}) and (\ref{ab}).
The wave numbers $k_L$ and $\kappa_L$ for $x\to-\infty$ are given by
\begin{eqnarray}
k_L&=&\sqrt{2m}\sqrt{E-(U_2-\mu)},\\
\kappa_L&=&\sqrt{2m}\sqrt{E+(U_2-\mu)}.
\end{eqnarray}
$k_L$ and $\kappa_L$ are propagating and localized waves for
$x\to -\infty$, which are obtained
by solving $E=\pm E^{\rm s}_p$ in terms of $p$.
We note that the localized $v$ component in Eq.~(\ref{step2asympt})
describes the (proximity) effect of the condensate in the normal region.

Using Eq.~(\ref{step2asympt}), we obtain the quasiparticle current
$J_{\rm q}$, as well as the energy flux of
quasiparticles $Q_{\rm q}$, in the normal region ($x\ll -\xi$) as
\begin{eqnarray}
J_{\rm q}=\frac{k_L}{mL}(1-|r|^2),\label{JqNS}\\
Q_{\rm q}=\frac{k_LE}{mL}(1-|r|^2).\label{QqNS}
\end{eqnarray}
Since the localized $v$ component in Eq.~(\ref{step2asympt}) does not give
rise to any contribution to the currents, 
$J_{\rm q}$ and $Q_{\rm q}$ reduce to those of free atoms which satisfy the relation $Q_{\rm
q}=EJ_{\rm q}$. From Eqs.~(\ref{asymptQ}) and (\ref{QqNS}), we obtain the transmission
(reflection) probability $W$ ($R$) as
\begin{eqnarray}
W&=&L\frac{k}{k_L}(a^2+b^2)|t|^2,\\
R&=&|r|^2.
\end{eqnarray}
Since the energy flux is conserved as shown in Appendix \ref{B}, they
satisfy the condition $R+W=1$.
We also obtain the ``transmission (reflection) probability'' $W_J$ ($R_J$) for
quasiparticle current from Eqs.~(\ref{asymptJq}) and (\ref{JqNS}), 
\begin{eqnarray}
W_J&=&\frac{k}{k_L}|t|^2=\frac{1}{L(a^2+b^2)}W,\label{WJNS}\\
R_J&=&|r|^2.
\end{eqnarray}
We again find that, $W$ and $W_J$ do not coincide with each other.
Because of $W_J<W$, we obtain $R_J+W_J<1$. This implies that
the quasiparticle current decreases on the superfluid region.
Since the ratio between $W$ and $W_J$ is given by
$L(a^2+b^2)=\sqrt{E^2+(gn_0)^2}/E$, $W$ and $W_J$ become equal when
$E/gn_0\gg 1$.

\begin{figure}
\centerline{\includegraphics{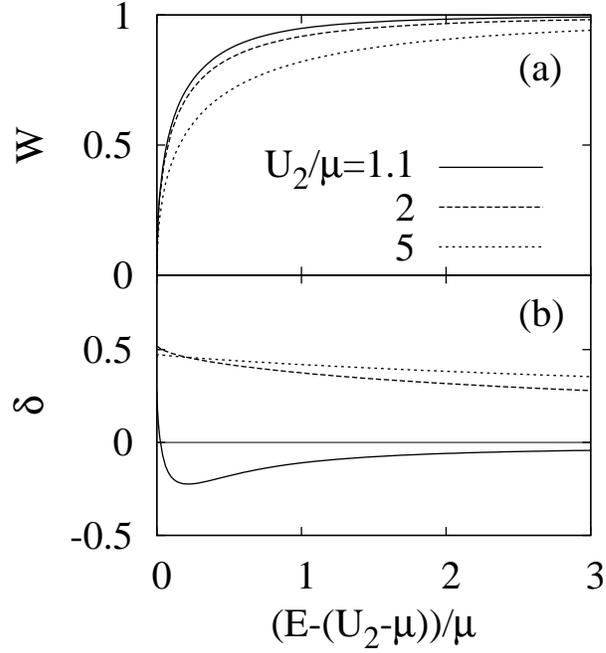}}
\caption{Transmission probability $W$ obtained from the energy flux
 (a) and phase shift $\delta$ (b) in both the N-S and S-N
 tunneling cases.}
\label{WqdeltaNS}
\end{figure}

\begin{figure}
\centerline{\includegraphics{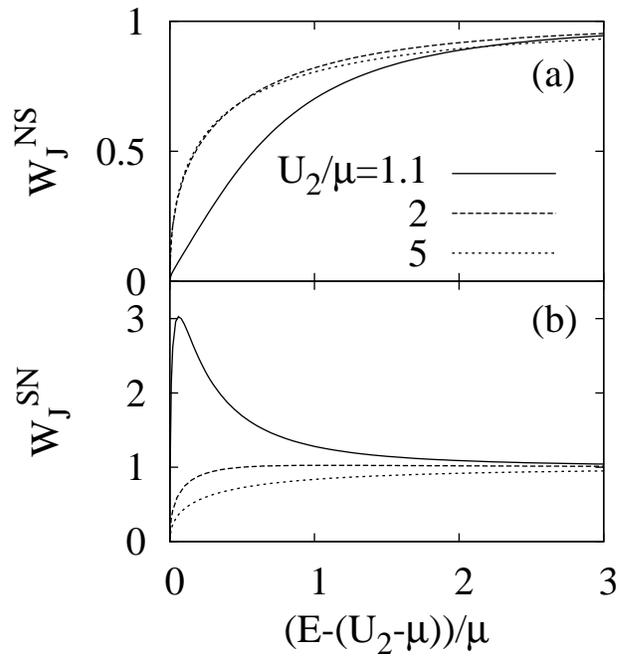}}
\caption{Transmission probabilities obtained from the quasiparticle current in the
 N-S ($W_J^{\rm NS}$) (a) and S-N ($W_J^{\rm
 SN}$) (b) tunneling cases. }
\label{WjNS}
\end{figure}

Figure~\ref{WqdeltaNS} shows the transmission probability $W$ obtained
from the energy flux and the phase shift $\delta\equiv{\rm arg}(t)$.
We also show the quasiparticle transmission probability $W_J^{\rm NS}$ in Fig.~\ref{WjNS} (a).
In Figs.~\ref{WqdeltaNS} and \ref{WjNS}, we note that the origin of $E$ is
taken to be $U_2-\mu$, because atoms are perfectly reflected when
$E<U_2-\mu$, leading to vanishing $W_J^{\rm NS}$ and $W$.
We find that both $W$ and $W_J$ decrease with decreasing $E$, and
$W$ and $W_J$ approach 0 when $E\to U_2-\mu$, while $\delta$
approaches a positive value in the limit of $E\to U_2-\mu$.
Thus, the anomalous tunneling behavior does not occur in the present case.

\begin{figure}
\centerline{\includegraphics{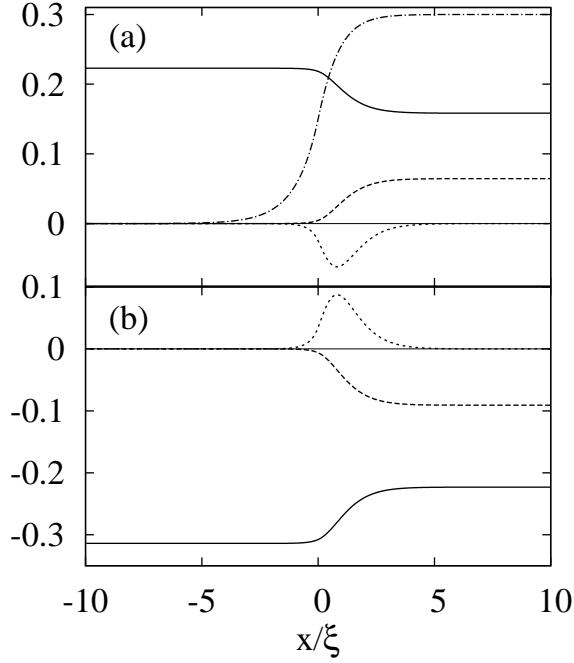}}
\caption{
 Spatial variation in quasiparticle current $J_{\rm q}(x)$
 (solid line), induced supercurrent $\Delta J_{\rm s}(x)$ (dashed line),
 and source term $S(x)$ (dotted line), when $U_2=2\mu$ and $E=1.01\mu$
 for the N-S tunneling (a) and S-N tunneling (b). Current density and
 source term are in units of $k_L/mL$ and $k_L/mL\xi$ in (a), and $k/mL$ and $k/mL\xi$
 in (b), respectively.
 The dash-dotted line in (a) indicates the condensate wave function
 $\Psi_0(x)$ in units of $\sqrt{n_0}/0.3$.
}
\label{JSNS}
\end{figure}

Figure~\ref{JSNS}(a) shows the spatial variation in quasiparticle current
$J_{\rm q}(x)$, source term $S(x)$ [defined by Eq.~(\ref{source})], as well as the induced supercurrent $\Delta
J_{\rm s}(x)$.
The existence of transmitted component of $J_{\rm q}(x)$ shows that the incident
current of free atoms from the normal region is converted into the
Bogoliubov excitations inside the condensate. 
Furthermore, one finds that $J_{\rm q}(x)$ decreases near the surface at
$x\simeq 0$, and the supercurrent $\Delta J_{\rm s}(x)$ is induced around the
surface. The source term $S(x)$ becomes negative near the surface of the
superfluid region reflecting the behaviors of $J_{\rm q}(x)$ and $\Delta
J_{\rm s}(x)$.
These phenomena indicate that injected atoms are Bose condensed in the
superfluid region, which give rise to the supercurrent $\Delta J_{\rm s}$. 
The condition $R_J+W_J<1$ reflects the
fact that a part of the incident current of free atoms is converted to
supercurrent inside the condensate.
The supercurrent $\Delta J_{\rm s}$ decreases as $E$ increases, because
the character of produced Bogoliubov phonon becomes close to that of
single-particle excitation, as $E$ increases.
As a result, $W_J$ approaches $W$ when $E\gg gn_0$.

We next consider the S-N tunneling.
Assuming that the incident Bogoliubov mode comes from $x=+\infty$,
the asymptotic solutions $(u,v)$ for $x\to\pm\infty$ are given by
\begin{eqnarray}
\begin{cases}
\left(\begin{array}{l}
u\\
v
\end{array}\right)=
t\left(
\begin{array}{l}
1\\
0
\end{array}\right)\frac{e^{-ik_Lx}}{\sqrt{L}}+
B\left(\begin{array}{l}
0\\
1
\end{array}
\right)\frac{e^{\kappa_L x}}{\sqrt{L}}, & (x\to-\infty),\\
\left(\begin{array}{l}
u\\
v
\end{array}\right)=
\left(
\begin{array}{l}
a\\
b
\end{array}
\right)e^{-ikx}+
r\left(
\begin{array}{l}
a\\
b
\end{array}
\right)e^{ikx}+
A\left(
\begin{array}{l}
-b\\
a\end{array}
\right)e^{-\kappa x}, & (x\to\infty).
\end{cases}
\label{asymptSN}
\end{eqnarray}
Using Eq.~(\ref{asymptSN}), the quasiparticle current $J_{\rm q}$ and
energy flux of quasiparticles $Q_{\rm q}$ in the normal region ($x\ll
-\xi$) are calculated as
\begin{eqnarray}
J_{\rm q}&=&\frac{-k_L}{mL}|t|^2,\label{JqSN}\\
Q_{\rm q}&=&\frac{-k_LE}{mL}|t|^2.\label{QqSN}
\end{eqnarray}
From Eqs.~(\ref{asymptQ}) and (\ref{QqSN}), we obtain the transmission (reflection)
probability $W$ ($R$) obtained from the energy flux as
\begin{eqnarray}
W&=&\frac{1}{L(a^2+b^2)}\frac{k_L}{k}|t|^2,\label{WQSN}\\
R&=&|r|^2.\label{RQSN}
\end{eqnarray}
Equations (\ref{WQSN}) and (\ref{RQSN}) again satisfy the condition $R+W=1$.
From Eqs.~(\ref{asymptJq}) and (\ref{JqSN}), we obtain the transmission
(reflection) probability $W_J$ ($R_J$) for quasiparticle current as
\begin{eqnarray}
W_J&=&\frac{k_L}{k}|t|^2=L(a^2+b^2)W,\label{WJSN}\\
R_J&=&|r|^2.
\end{eqnarray}
In contrast to the N-S tunneling, it is clear from Eq.~(\ref{WJSN}) that
$W_J>W$, which leads to the condition $R_J+W_J>1$.
This implies that the quasiparticle current is supplied around the
surface at $x\simeq 0$. When $E\gg gn_0$, $W_J$ reduces to $W$.

It can be generally shown for the Bogoliubov coupled Eqs.~(\ref{Bogoliubovu}) and (\ref{Bogoliubovv}) that
$W$ and $\delta$ are both independent of whether the incident wave comes
from $x=-\infty$ (N-S tunneling) or $x=+\infty$ (S-N tunneling)
\cite{Tsuchiya2}. 
Hence, $W$ and $\delta$ in the S-N tunneling case are the same as those
in the case of N-S tunneling in Fig.~\ref{WqdeltaNS}.

The transmission probability for quasiparticle current $W_J^{\rm SN}$ is
shown in Fig.~\ref{WjNS}(b).  
We find that $W_J^{\rm SN}$ is enhanced to be greater than unity at low
energies, due to the factor $L(a^2+b^2)=\sqrt{E^2+(gn_0)^2}/E$ in Eq.~(\ref{WJSN}).

When the Bogoliubov phonons propagate toward the S-N phase boundary,
Fig.~\ref{JSNS}(b) shows that atoms evaporate from the surface.
We also find that $J_{\rm q}(x)$ changes near the surface and the
supercurrent $\Delta J_{\rm s}$ is induced, which flows toward the boundary.
$S(x)$ becomes positive around the surface of the superfluid region,
reflecting the behavior of $J_{\rm q}(x)$ and $\Delta J_{\rm s}(x)$.
This supercurrent $\Delta J_{\rm s}(x)$ is considered to be induced by the reflected
Bogoliubov mode, which twists the condensate phase in the superfluid
region.
The fact of $R_J+W_J>1$ reflects that $J_{\rm q}(x)$ increases during
the tunneling through the S-N phase boundary.

\section{conclusions}
\label{sec6}

To summarize, we have investigated tunneling effects of Bogoliubov
excitations at $T=0$.
We have extended our previous work to the case when the condensate
densities are different on the left and right of the barrier. 
Within the frame work of the Bogoliubov theory, we have evaluated
the transmission probability, phase shift as well as the energy flux and
quasiparticle current carried by Bogoliubov excitations.
We showed that, while the energy flux is conserved, the quasiparticle
current is not conserved. The excess quasiparticle current is actually
cancelled out by the counterflow of supercurrent, which is induced by
the back-reaction effects of Bogoliubov phonons on the condensate.
In the case of a rectangular potential barrier,
we directly showed that the induced supercurrent satisfies the Josephson
relation with respect to the twisted phase by Bogoliubov phonons.
When the condensate has different densities on the left and right of the
barrier, the supercurrent is induced in the region far from the barrier potential.
We also studied the tunneling of atoms from the normal region to the
superfluid region, as well as the tunneling of excitations from the
superfluid region to the normal region.
In the former case, we showed that supercurrent is induced inside a
condensate by injecting free atoms from outside. In the latter case, we
found that atoms evaporate from the superfluid-normal state phase
boundary, when Bogoliubov excitations propagate toward the surface of
the superfluid region.
We think these results can be of interest for the investigation of
Bogoliubov mode and its connection to the superfluidity of BECs in
ultracold atomic gases.

\acknowledgments

We wish to thank I. Danshita, S. Watabe, D. Takahashi, K. Kamide, N. Yokoshi, S. Inouye,
F. Dalfovo, S. Kurihara, and Y. Kato for stimulating discussions.
We acknowledge M. Machida, T. Suzuki, M. Ueda, and T. Nikuni for valuable comments.
This work was supported by a Grant-Aid for Scientific Research from
MEXT, Japan and the CTC program of Japan.

\appendix
\section{Formalism of weakly interacting Bose gases}
\label{A}
In this appendix, we summarize the formalism of weakly interacting Bose
gases developed in \cite{Bogoliubov, Gross, Pitaevskii2, Fetter, Griffin}. 
We introduce approximations for inhomogeneous Bose condensates including
the Bogoliubov approximation used in this paper.

We consider an interacting Bose gas described by the Hamiltonian,
\begin{equation}
\hat K=\int d\bm r\ \hat\psi^\dagger(\bm r)\left(-\frac{\nabla^2}{2m}+U(\bm
					r)-\mu\right)\hat\psi(\bm r)+\frac{g}{2}\int d\bm r\ \hat\psi^\dagger(\bm r)\hat\psi^\dagger(\bm r)\hat\psi(\bm r)\hat\psi(\bm r),
\label{Hamiltonian}
\end{equation}
where $\hat\psi(\bm r)$ is the Bose field operator, $\mu$ is the
chemical potential, and $U(\bm r)$ is an external potential.
We assume a contact interaction between atoms $g\delta(\bm r-\bm
r^\prime)$ with the coupling constant $g=4\pi a_s/m$, where $a_s>0$ is the
$s$-wave scattering length.

In the Bose condensed phase, 
we divide the field operator $\hat\psi(\bm r)$ into the sum of the condensate
wave function $\Psi_0(\bm r)=\langle\hat\psi(\bm r)\rangle$ and the fluctuation part $\delta\hat\psi$ as 
\begin{equation}
\hat\psi(\bm r)=\Psi_0(\bm r)+\delta\hat\psi(\bm r).
\label{fieldop}
\end{equation}
Substituting Eq.~(\ref{fieldop}) into Eq.~(\ref{Hamiltonian}),
we approximately evaluate the cubic and quartic terms with respect to
$\delta\hat\psi$ and $\delta\hat\psi^\dagger$ as
\begin{eqnarray}
\delta\hat\psi^\dagger \delta\hat\psi \delta\hat\psi&\simeq&
 2\tilde{n}\delta\hat\psi+\tilde{m}\delta\hat\psi^\dagger,\\
\delta\hat\psi^\dagger\delta\hat\psi^\dagger\delta\hat\psi\delta\hat\psi
&\simeq& 4\tilde{n}\delta\hat\psi^\dagger\delta\hat\psi+\tilde{m}^\ast\delta\hat\psi\delta\hat\psi+\tilde{m}\delta\hat\psi^\dagger\delta\hat\psi^\dagger,
\end{eqnarray}
where $\tilde n(\bm r)=\langle\delta\hat\psi^\dagger\delta\hat\psi\rangle$ is
the non-condensate density, and $\tilde
m(\bm r)=\langle\delta\hat\psi\delta\hat\psi\rangle$ is the so-called anomalous
average \cite{Griffin}. 
In this mean-field approximation, Eq.~(\ref{Hamiltonian}) reduces to
\begin{eqnarray}
\hat K&=&\hat K_0+\hat K_1+\hat K_2,\\
\hat K_0&=&\int d\bm r\ \Psi_0^\ast\hat T\Psi_0+\frac{g}{2}\int d\bm r\ |\Psi_0|^2,\\
\hat K_1&=&\int d{\bm r}\left[\left(\hat T\Psi_0+g(|\Psi_0|^2+2\tilde{n})\Psi_0+g\tilde{m}\Psi_0^\ast\right)\delta\hat\psi^\dagger + {\rm h.c.}\right],\\
\hat K_2&=&\int d{\bm r}\left[\delta\hat\psi^\dagger\left(\hat T+2g(|\Psi_0|^2+\tilde{n})\right)\delta\hat\psi+\frac{g}{2}\left((\Psi_0^2+\tilde{m})\delta\hat\psi^\dagger\delta\hat\psi^\dagger+{\rm h.c.}\right)\right].\label{quadratic}
\end{eqnarray}
Here, $\hat T\equiv -\frac{\nabla^2}{2m}+U(\bm r)-\mu$.
From the condition that the linear term in terms of $\delta\psi$ and
$\delta\psi^\dagger$ vanishes, we obtain the (generalized) Gross-Pitaevskii equation \cite{Pitaevskii2,Gross}, 
\begin{eqnarray}
\left[-\frac{\nabla^2}{2m}+U(\bm r)+g(|\Psi_0|^2+2\tilde
 n)\right]\Psi_0+g\tilde m\Psi_0^\ast=\mu \Psi_0\ .
\label{gGP1}
\end{eqnarray}
The quadratic term $\hat K_2$ in Eq.~(\ref{quadratic}) can be diagonalized by the Bogoliubov transformation \cite{Fetter}
\begin{eqnarray}
\delta\hat\psi(\bm r)=\sum_j\left[u_j(\bm r)\hat\alpha_j-v_j(\bm r)^\ast\hat\alpha_j^\dagger\right],\label{excitation}\\
\delta\hat\psi^\dagger(\bm r)=\sum_j\left[u_j(\bm r)^\ast\hat\alpha_j^\dagger-v_j(\bm r)\hat\alpha_j\right],
\end{eqnarray}
where $\hat\alpha_j^\dagger$ is the creation operator of a Bogoliubov excitation in the $j$th state, which obeys the bosonic
commutation relations,
\begin{equation}
[\hat\alpha_i,\hat\alpha_j^\dagger]=\delta_{i,j},\ 
[\hat\alpha_i,\hat\alpha_j]=[\hat\alpha_i^\dagger,\hat\alpha_j^\dagger]=0.
\end{equation}
Diagonalization of $\hat K_2$ is achieved when $(u_j(\bm r),v_j(\bm r))$
satisfy the following generalized Bogoliubov equations \cite{Pitaevskii2,Fetter,Griffin}:
\begin{eqnarray}
\left[-\frac{\nabla^2}{2m}+U(\bm
 r)+2g(|\Psi_0|^2+\tilde{n})-\mu\right]u_j&\!-\!&g(\Psi_0^2+\tilde{m})v_j=E_ju_j,\label{gBogoliubovu}\\
\left[-\frac{\nabla^2}{2m}+U(\bm r)+2g(|\Psi_0|^2+\tilde{n})-\mu\right]v_j&\!-\!&g((\Psi_0^\ast)^2+\tilde{m}^\ast)u_j=-E_jv_j.
\label{gBogoliubovv}
\end{eqnarray}
Then, we have
\begin{equation}
\hat K=\hat K_0+\sum_j E_j\hat\alpha_j^\dagger\hat\alpha_j-\sum_j
 E_j\int d\bm r\ |v_j|^2\ .
\label{diaenergy}
\end{equation}
The last term in Eq.~(\ref{diaenergy}) is the so-called quantum
depletion, describing the non-condensate due to the
repulsive interaction between atoms.
It remains finite even at $T=0$, where $\langle\hat\alpha_j^\dagger\hat\alpha_j\rangle=0$.

We note that Eq.~(\ref{gGP1}) involves terms originating from excitations
($2g\tilde n\Psi_0$ and $g\tilde m\Psi_0^\ast$). This reflects the
fact that the condensate and excitations affect each other.
In the Bogoliubov approximation, both $\tilde{n}$ and $\tilde{m}$ are
neglected, so that effects of Bogoliubov excitations on the condensate
are not taken into account.  
Equation~(\ref{gGP1}) without $\tilde{n}$ and $\tilde{m}$ is the ordinary
static GP equation \cite{Pitaevskii2, Gross}, while
Eqs.~(\ref{gBogoliubovu}) and (\ref{gBogoliubovv}) without $\tilde{m}$ and
$\tilde{n}$ are the Bogoliubov coupled equations.
The Bogoliubov approximation is valid at very low temperatures where
$\tilde n$ and $\tilde m$ are very small.

The approximation keeping both $\tilde{n}$ and $\tilde{m}$ is the
Hartree-Fock-Bogoliubov approximation \cite{Griffin}. This approximation is 
valid at finite temperatures where non-condensate fluctuation cannot
be neglected. In this approximation, however, the excitation spectrum in
a uniform system has an energy gap \cite{Griffin}. 
This is inconsistent with the Hugenholtz-Pines theorem
\cite{Hugenholtz}, which states that the excitation 
spectrum must be gapless in the BEC phase.
Keeping $\tilde{n}$ but neglecting $\tilde{m}$ is referred to as the
Popov approximation \cite{Pitaevskii,Griffin}. 
This approximation is also considered to be valid at finite temperatures.
Since it yields a gapless excitation spectrum, it has been widely used in the
study of BEC at finite temperatures \cite{Pitaevskii, Griffin,Hutchinson}. 
Tunneling properties of Bogoliubov excitations at finite temperatures
have been also studied using the Popov approximation \cite{Kato}.

\section{Conservation laws for Bogoliubov excitations}
\label{B}

In this appendix, we discuss the conservation laws in terms of
quasiparticle current and energy flux associated with Bogoliubov
excitations.
In the Bogoliubov mean-field approximation, the total number density
$n\equiv\langle\hat\psi^\dagger\hat\psi\rangle$ and 
total current density ${\bm J}\equiv(1/m){\rm
Im}\langle\hat\psi^\dagger\nabla\hat\psi\rangle$ are, respectively, given by
\begin{eqnarray}
n&=&n_{\rm s}+\sum_j\left(n_{u_j}+n_{v_j}\right)\langle\hat\alpha_j^\dagger\hat\alpha_j\rangle+\sum_jn_{v_j},\label{density}
\\
{\bm J}&=&{\bm J}_{\rm s}+\sum_j\left({\bm J}_{u_j} - {\bm J}_{v_j}\right)
\langle\hat\alpha_j^\dagger\hat\alpha_j\rangle-\sum_j {\bm J}_{v_j}.
\label{current}
\end{eqnarray}
In this appendix, the index for eigenstates $j$ is explicitly written.
Note that in our tunneling problem of Bogoliubov excitation, each eigenstate
in Eqs.~(\ref{density}) and (\ref{current}) corresponds to
Bogoliubov excitation with energy $E$ injected from $x=-\infty$ or $x=\infty$.
Here, $n_{\rm s}\equiv |\Psi_0|^2$ describes the condensate density and 
\begin{equation}
{\bm J}_{\rm s}=\frac{1}{m}{\rm Im}(\Psi_0^\ast\nabla \Psi_0)
\label{Js}
\end{equation}
is the supercurrent density carried by the condensate.
$n_{u_j}$, $n_{v_j}$, ${\bm J}_{u_j}$, and ${\bm J}_{v_j}$ are,
respectively, given by
\begin{eqnarray}
n_{u_j}&=&|u_j|^2,\\
n_{v_j}&=&|v_j|^2,\\
{\bm J}_{u_j}&=&\frac{1}{m}{\rm Im}(u_j^\ast\nabla u_j),\\
{\bm J}_{v_j}&=&\frac{1}{m}{\rm Im}(v_j^\ast\nabla v_j).
\end{eqnarray}
The total number density $n$ and total current density $\bm J$ satisfy the continuity equation
\begin{equation}
\partial_t n+\nabla\cdot \bm J=0. 
\label{totaldensity}
\end{equation}
Since the second terms in Eqs.~(\ref{density}) and (\ref{current})
describe the quasiparticle contributions, the quasiparticle density
$n_{{\rm q},j}$ and quasiparticle current ${\bm J}_{{\rm q},j}$ are, respectively, given by 
\begin{eqnarray}
n_{{\rm q},j}=n_{u_j}+n_{v_j},\label{qdensity}\\
{\bm J}_{{\rm q},j}={\bm J}_{u_j}-{\bm J}_{v_j}.\label{qcurrent}
\end{eqnarray} 
Equations~(\ref{qdensity}) and (\ref{qcurrent}) show that both the quasiparticle
density $n_{{\rm q},j}$ and current ${\bm J}_{{\rm q},j}$ consist of two
components originating from $u_j$ and $v_j$.
We note that the current density of $v$component appears as $-{\bm J}_{v_j}$ in Eq.~(\ref{qcurrent}). 
Thus, in a uniform system, a Bogoliubov phonon is accompanied by two
current components, ${\bm J}_{u_j}=({\bm p}/m)a^2$ and $-{\bm J}_{v_j}=-({\bm
p}/m)b^2$, where $a$ and $b$ are given in Eq.~(\ref{ab}).
Hence, the $v$ component flows in the oppose direction to the $u$ component.
Indeed, these counterpropagating currents were recently observed \cite{Vogels}.
The last terms in Eqs.~(\ref{density}) and (\ref{current}) describe
effects of quantum depletion.

To derive the continuity equation for quasiparticles, it is convenient to use the
time-dependent Bogoliubov equations \cite{Castin} for $\left(u({\bm r},t),v({\bm
r},t)\right)$,
\begin{eqnarray}
i\tau_3\partial_t
\left(\begin{array}{l}
u\\
v
\end{array}\right)=
\left(
\begin{matrix}
\hat{h} & -g\Psi_0^2\\
-g(\Psi_0^\ast)^2 & \hat{h}
\end{matrix}
\right)
\left(
\begin{array}{l}
u\\
v
\end{array}
\right),
\label{TDBogoliubov}
\end{eqnarray}
where $\hat h\equiv -\frac{\nabla^2}{2m}+U(\bm r)+2g|\Psi_0|^2-\mu$.
Equation~(\ref{TDBogoliubov}) reduces to Eqs.~(\ref{Bogoliubovu}) and
(\ref{Bogoliubovv}) in the
stationary state, $\left(u({\bm r},t),v({\bm
r},t)\right)=e^{-iE_jt}(u_j({\bm r}),v_j(\bm r))$. 

Using Eq.~(\ref{TDBogoliubov}), one obtains the continuity equations for
$u_j$ and $v_j$, as
\begin{eqnarray}
\partial_tn_{u_j}+\nabla\cdot{\bm J}_{u_j}=\frac{S_j}{2},\label{contu}\\
\partial_t n_{v_j}-\nabla\cdot{\bm J}_{v_j}=\frac{S_j}{2}\label{contv},
\end{eqnarray}
where
\begin{equation}
S_j=-4g{\rm Im}\left(\Psi_0^2u_j^\ast v_j \right).
\label{source}
\end{equation}
Thus, the continuity equation for quasiparticles is given by
\begin{eqnarray}
\partial_t n_{{\rm q},j}+\nabla\cdot{\bm J}_{{\rm q},j}=S_j\ .
\label{qcontinuity}
\end{eqnarray}
In Eq.~(\ref{qcontinuity}), $S_j$ works as a source term. This means
that the total number of quasiparticles is {\it not conserved} when
$S_j\neq 0$. In a uniform system, one finds $S_j=0$, so that the number of
quasiparticles is conserved. On the other hand, since the source term
$S_j$ is finite near the potential barrier in our 
tunneling problem, the number of quasiparticles is {\it not}
conserved.

We next consider the energy flux. For this purpose, we define the energy
density operator $\hat\rho$ as
\begin{equation}
\hat\rho=\hat\psi^\dagger\hat T\hat\psi+\frac{g}{2}\hat\psi^\dagger\hat\psi^\dagger\hat\psi\hat\psi,
\end{equation}
where $\hat T$ is defined below Eq.~(\ref{quadratic}). Using the Heisenberg equation
$i\partial_t{\hat\psi}=\hat T\hat\psi+g\hat\psi^\dagger\hat\psi\hat\psi,$
one obtains the continuity equation for energy density $\hat\rho$ as,
\begin{equation}
\partial_t\hat\rho+\nabla\cdot\hat{\bm Q}=0.
\end{equation}
Here, $\hat{\bm Q}$ is the energy flux operator, defined by
\begin{eqnarray}
\hat{\bm Q}&=&\frac{i}{2m}\left[(\nabla\hat\psi^\dagger)\left(\hat
						       T\hat\psi+g\hat\psi^\dagger\hat\psi\hat\psi\right)-{\rm
h.c.}\right]\nonumber\\
&=&-\frac{1}{m}{\rm Re}\left[(\nabla\hat\psi^\dagger)(\partial_t{\hat\psi})\right].
\end{eqnarray}
Substituting Eq.~(\ref{fieldop}) into $\rho=\langle\hat\rho\rangle$ and
retaining terms up to $O(\delta\hat\psi^2)$, we obtain
\begin{eqnarray}
\rho&=&\Psi_0^\ast\hat T\Psi_0+\frac{g}{2}|\Psi_0|^4
+\langle\delta\hat\psi\hat T\delta\hat\psi\rangle\nonumber\\
&&+\frac{g}{2}\left(\Psi_0^2\langle\delta\hat\psi^\dagger\delta\hat\psi^\dagger\rangle+4|\Psi_0|^2\langle\delta\hat\psi^\dagger\delta\hat\psi\rangle+(\Psi_0^\ast)^2\langle\delta\hat\psi\delta\hat\psi\rangle\right).\label{rho}
\end{eqnarray}
In obtaining Eq.~(\ref{rho}), we have used $\langle\delta\hat\psi\rangle=0$.
Using Eqs.~(\ref{excitation}), (\ref{gBogoliubovu}), and
(\ref{gBogoliubovv}), we obtain 
\begin{eqnarray}
\rho=\rho_0+\sum_jE_j\left(n_{u_j}-n_{v_j}\right)\langle\hat\alpha_j^\dagger\hat\alpha_j\rangle-\sum_jE_jn_{v_j}-\frac{i}{4}\sum_jS_j,
\label{energyd}
\end{eqnarray}
where 
\begin{equation}
\rho_0=\Psi_0^\ast\hat T\Psi_0+\frac{g}{2}|\Psi_0|^4
\end{equation}
is the condensate energy density. 
Since the energy density $\rho$ is a real quantity, the last term in
Eq.~(\ref{energyd}) must vanish, which gives
\begin{equation}
\sum_j S_j=0.
\label{source0}
\end{equation}

The energy flux $\bm Q\equiv \langle\hat{\bm Q}\rangle$ can be also calculated
in the same manner. The result is
\begin{equation}
\bm Q=\bm Q_0+\sum_j E_j(\bm J_{u_j}+\bm
 J_{v_j})\langle\hat\alpha_j^\dagger\hat\alpha_j\rangle+\sum_jE_j\bm
 J_{v_j}.
\label{energyf}
\end{equation}
Here,
\begin{equation}
\bm Q_0=\frac{i}{2m}\left[\left(\hat T\Psi_0+g(|\Psi_0|^2+2\tilde
			   n)\Psi_0+g\tilde
			   m\Psi_0^\ast\right)(\nabla\Psi_0^\ast)-{\rm c.c.}\right]
\end{equation}
is interpreted as the energy flux carried by the condensate.
Actually, $\bm Q_0$ identically vanishes when $\Psi_0$ satisfies the (generalized) GP equation.

The second terms in Eqs.~(\ref{energyd}) and (\ref{energyf})
describe the quasiparticle contributions. Thus, the energy density
for quasiparticles $\rho_{{\rm q},j}$ and energy flux for quasiparticles $\bm
Q_{{\rm q},j}$ are, respectively, given by
\begin{eqnarray}
\rho_{{\rm q},j}=E_j(n_{u_j}-n_{v_j}),\label{energydq}\\
\bm Q_{{\rm q},j}=E_j(\bm J_{u_j}+\bm J_{v_j}).\label{energyfq}
\end{eqnarray}
Equation~(\ref{energydq}) shows that the $v$ component has a negative energy
density $-E_jn_{v_j}$. In Eq.~(\ref{energyfq}), the $v$ component
appears as $+E_j{\bm J}_{v_j}$, which is in contrast to $\bm J_{{\rm
q},j}$ in Eq.~(\ref{qcurrent}), where the $v$ component appears as $-\bm J_{v_j}$.
This is because the $v$ component has a negative energy $-E_j$ and counterpropagating current density $-\bm J_{v_j}$.
In contrast to the nonconserved quasiparticle number density in Eq.~(\ref{qcontinuity}), the continuity
equation with respect to the energy density has no source term, as 
\begin{equation}
\partial_t\rho_{{\rm q},j}+\nabla\cdot {\bm Q}_{{\rm q},j}=0.
\label{econserv}
\end{equation}
Namely, ${\bm Q}_{{\rm q},j}$ is conserved everywhere in the stationary state.

To examine the origin of the source term $S_j$ in Eq.~(\ref{qcontinuity}), it is
convenient to consider the divergence of Eq.~(\ref{current}) in the stationary state,
\begin{equation}
\nabla\cdot{\bm J}=\nabla\cdot{\bm J}_{\rm s}+\sum_j
 S_j\langle\hat\alpha_j^\dagger\hat\alpha_j\rangle.
\label{divcurrent}
\end{equation}
In obtaining Eq.~(\ref{divcurrent}), we have used
Eqs.~(\ref{contu})-(\ref{source}) and (\ref{source0}).
The static GP Eq.~(\ref{GP}) guarantees the conservation of the
supercurrent ($\nabla\cdot \bm J_{\rm s}=0$), so that
Eq.~(\ref{divcurrent}) contradicts with the conservation of the total
current $\bm J$ obtained from Eq.~(\ref{totaldensity}), unless the last
term in Eq.~(\ref{divcurrent}) vanishes identically.

This inconsistency arises because effects of quasiparticles on
condensates (back-reaction effect) are completely neglected in the Bogoliubov
approximation. This problem can be solved by including quasiparticle
contribution to the condensate in the GP equation as
\begin{eqnarray}
\left(-\frac{\nabla^2}{2m}+U({\bm r})+g|\Psi_0|^2\right)\Psi_0-2g\sum_j
 u_jv_j^\ast\langle\hat\alpha_j^\dagger\hat\alpha_j\rangle\Psi_0^\ast=\mu\Psi_0\ .
\label{gGP}
\end{eqnarray}
In this modified GP equation, the last term on the left-hand side
originates from the anomalous average $\tilde m$ in Eq.~(\ref{gGP1}) as
\begin{equation}
g\langle\delta\hat\psi\delta\hat\psi\rangle\Psi_0^\ast=-2g\sum_ju_jv_j^\ast\langle\hat\alpha_j^\dagger\hat\alpha_j\rangle\Psi_0^\ast-g\sum_ju_jv_j^\ast\Psi_0^\ast.
\end{equation}
Using Eq.~(\ref{gGP}), the conservation of the supercurrent
($\nabla\cdot\bm J_{\rm s}=0$) is modified to be
\begin{equation}
\nabla\cdot{\bm J}_{\rm
 s}=-\sum_jS_j\langle\hat\alpha_j^\dagger\hat\alpha_j\rangle.
\label{contJs}
\end{equation}
Substituting Eq.~(\ref{contJs}) into Eq.~(\ref{divcurrent}), we obtain
the expected conservation of the total current $\nabla\cdot{\bm J}=0$.
In the case of the one-dimensional model we are using in this paper, when
we integrate Eq.~(\ref{contJs}) in terms of $x$ from $-\infty$ to $x$,
we obtain Eq.~(\ref{deltaJs2}).

\end{document}